\documentclass[paper,11pt]{JHEP}
\usepackage[centertags]{amsmath}
\usepackage{amsfonts} 
\usepackage{amssymb} 
\usepackage{amsthm}
\usepackage{graphicx}
\usepackage{bm}
\newcommand{\be}{\begin{equation}}
\newcommand{\eeq}{\end{equation}}
\newcommand{\bea}{\begin{eqnarray}}
\newcommand{\eea}{\end{eqnarray}}
\newcommand{\ba}{\begin{array}}
\newcommand{\ea}{\end{array}}

\newcommand{\eq}[1]{Eq.~(\ref{#1})}
\newcommand{\comm}[2]{\left[#1,#2\right]}
\newcommand{\ii}{\mathrm{i}}
\newcommand{\ee}{\mathrm{e}}

\newcommand{\Tr}{\mathrm{Tr}\,}
\newcommand{\tr}{\mathrm{tr}\,}
\newcommand{\diag}{\mathrm{diag}\,}

\newcommand{\cC}{{\mathcal{C}}}

\newcommand{\cE}{{\mathcal{E}}}
\newcommand{\cF}{{\mathcal{F}}}

\newcommand{\cM}{{\mathcal{M}}}
\newcommand{\cN}{{\mathcal{N}}}
\newcommand{\cP}{{\mathcal{P}}}
\newcommand{\cQ}{{\mathcal{Q}}}
\newcommand{\cR}{{\mathcal{R}}}

\newcommand{\cZ}{{\mathcal{Z}}}

\newcommand{\im}{\mbox{Im}\,}
\newcommand{\one}{{\rm 1\kern -.9mm l}}

\newcommand{\Pf}{\mathrm{Pf}}
\newcommand{\ve}[1]{{\vec e}_{#1}}
\newdimen\tableauside\tableauside=1.0ex
\newdimen\tableaurule\tableaurule=0.4pt
\newdimen\tableaustep

\def\phantomhrule#1{\hbox{\vbox to0pt{\hrule height\tableaurule
width#1\vss}}}
\def\phantomvrule#1{\vbox{\hbox to0pt{\vrule width\tableaurule
height#1\hss}}}
\def\sqr{\vbox{%
 \phantomhrule\tableaustep
\hbox{\phantomvrule\tableaustep\kern\tableaustep\phantomvrule\tableaustep}%
 \hbox{\vbox{\phantomhrule\tableauside}\kern-\tableaurule}}}
\def\squares#1{\hbox{\count0=#1\noindent\loop\sqr
 \advance\count0 by-1 \ifnum\count0>0\repeat}}
\def\tableau#1{\vcenter{\offinterlineskip
 \tableaustep=\tableauside\advance\tableaustep by-\tableaurule
 \kern\normallineskip\hbox
   {\kern\normallineskip\vbox
     {\gettableau#1 0 }%
    \kern\normallineskip\kern\tableaurule}%
 \kern\normallineskip\kern\tableaurule}}
\def\gettableau#1 {\ifnum#1=0\let\next=\null\else
 \squares{#1}\let\next=\gettableau\fi\next}
\newcommand{\Yfund}{\tableau{1}}
\newcommand{\Ysymm}{\tableau{2}}
\newcommand{\Yasymm}{\tableau{1 1}}

\title{Stringy instanton effects in $\mathcal N=2$ gauge theories}
\author{
Hossein Ghorbani, Daniele Musso\\
Dipartimento di Fisica Teorica, Universit\`a di Torino\\
and I.N.F.N. - sezione di Torino \\
Via P. Giuria 1, I-10125 Torino, Italy\\
E-mail:\email{ ghorbani@to.infn.it, mussod@to.infn.it}
}

\author{
Alberto Lerda\\
Dipartimento di Scienze e Tecnologie Avanzate, Universit\`a del Piemonte
Orientale\\
and I.N.F.N. - Gruppo Collegato di Alessandria - sezione di Torino\\
Viale T. Michel  11, I-15121 Alessandria, Italy\\
E-mail:\email{ lerda@to.infn.it}
}

\abstract{We study the non-perturbative effects induced by stringy instantons 
on $\cN=2$ $\mathrm{SU}(N)$ gauge
theories in four dimensions, realized on fractional D3 branes in a $\mathbb{C}^3/\mathbb{Z}_3$
orientifold. The stringy instantons, corresponding to D(--1) branes that occupy a node
of the orientifold quiver diagram where no D3 brane is present, have the right content of
zero-modes to produce non-perturbative terms in the four-dimensional effective
action. In the SU(2) theory these terms have the same structure for all instanton
numbers and yield a series of non-perturbative corrections to the prepotential.
We explicitly compute these corrections up to instanton number $k=5$ using localization
methods. 
}
\keywords{Superstrings, D-branes, Gauge Theories, Instantons}
\preprint{DFTT/24/2010}

\begin{document}

\section{Introduction and motivations}
\label{sec:intro}
The study of the non-perturbative regime of supersymmetric gauge theories has
always attracted great interest (for reviews see, for example, \cite{Shifman:1999mv}-\nocite{Dorey:2002ik}\cite{Bianchi:2007ft}). 
In the last decade remarkable progress in this field 
has been achieved using string inspired methods, {\it i.e.} realizing the gauge theories on the world-volume of space-filling D-branes embedded in supersymmetric string compactifications and 
introducing the non-perturbative corrections by means of localized branes, like 
D-instantons or totally wrapped Euclidean 
branes \cite{Witten:1995gx}-\nocite{Douglas:1995bn,Douglas:1996uz,Green:1997tn,
Green:2000ke}\cite{Billo:2002hm} 
(for a recent review see, for instance, \cite{Blumenhagen:2009qh}). 
This stringy setup has allowed
to reproduce in a nice and unified framework many different features and results 
of the standard instanton calculus for supersymmetric gauge theories, 
like for instance the ADHM construction \cite{Atiyah:1978ri}, the classical instanton profile
and the non-perturbative corrections to prepotentials or superpotentials
in various models. 

On the other hand, the observation that instantons can be described 
as branes within branes has paved the
way to several interesting generalizations corresponding to instanton
configurations that do not 
admit a standard gauge theory interpretation but still have a natural realization in terms of 
D-branes. We shall refer to this type of configurations as ``exotic'' or ``stringy'' instantons
as opposed to the ``ordinary'' gauge instantons. The latter correspond to localized
branes that share with the space-filling branes all features except their dimensionality.
In the simplest setups where the four-dimensional gauge theory is engineered with D3 branes, 
the ordinary instantons are described by D(--1) branes of the same kind, 
while in more general string compactifications where the gauge sector is realized 
on D$(3 + p)$ branes wrapped on a $p$-cycle $\cC$, the ordinary instantons correspond 
to Euclidean D$(p-1)$ branes totally wrapped on $\cC$. Different types of D(--1) branes (for example
with different Chan-Paton structures), or Euclidean branes wrapped on cycles $\cC'
\not=\cC$ correspond, instead, to stringy instantons that do not have a clear field-theory
interpretation, at least from a four-dimensional point of view%
\footnote{Some stringy instantons configurations have a nice field-theory
interpretation in eight dimensions as shown in \cite{Billo':2009gc}.}. Despite this fact,
or maybe precisely for this fact, the stringy instantons have recently attracted much interest
since they can generate novel types of interactions which are perturbatively forbidden and
whose strength is not linked to the gauge theory scale. This feature is very welcome
in the search for semi-realistic string scenarios for the physics beyond the Standard Model 
where a hierarchy between various Majorana masses and Yukawa couplings is expected.
Indeed, in some specific contexts the stringy instantons have been indicated as possible 
sources of neutrino masses \cite{Blumenhagen:2006xt}-\nocite{Ibanez:2006da}\cite{Ibanez:2007rs},
of certain Yukawa couplings in GUT models \cite{Blumenhagen:2007zk}, or of non-perturbative
contributions that may be relevant for moduli stabilization \cite{Camara:2007dy,Blumenhagen:2007sm}.
Other interesting applications of stringy instantons 
can be found in \cite{Argurio:2007vqa}-\nocite{Bianchi:2007wy,Blumenhagen:2007bn,Cvetic:2007qj,Ibanez:2007tu,Petersson:2007sc,Bianchi:2007rb,Cvetic:2008ws,Cvetic:2008hi,
Forcella:2008au,Billo':2008sp,Billo':2008pg,Uranga:2008nh,Ferretti:2009tz,Angelantonj:2009yj,Bianchi:2009bg,Cvetic:2009yh,Petersson:2010qu}\cite{Blumenhagen:2010dt}.

{From} a conformal field theory point of view, in the ordinary gauge instanton configurations the
mixed open strings suspended between the instantonic and the space-filling
branes have four directions with mixed Neumann-Dirichlet (ND) boundary conditions, and
possess massless excitations in the Neveu-Schwarz sector which
describe the size and gauge orientation of field theoretical instanton solutions. On the other
hand, in the exotic cases the mixed open strings either have extra twisted directions besides 
the four ND space-time directions, or are characterized by different types of Chan-Paton factors
at their end-points.
As a consequence, the bosonic moduli corresponding 
to the instanton size are missing and certain fermionic zero-modes become difficult to saturate. 
These unwanted fermionic zero-modes must be either removed by appropriate orientifold projections 
\cite{Argurio:2007vqa,Bianchi:2007wy}, or lifted with fluxes \cite{Blumenhagen:2007bn,Billo':2008sp,Billo':2008pg} 
or with other mechanisms \cite{Petersson:2007sc,Ferretti:2009tz}. 

Parallel to these developments, the application of
localization techniques to the computation of the instanton partition functions, originally
pioneered by N. Nekrasov \cite{Nekrasov:2002qd}-\nocite{Losev:2003py}\cite{Nekrasov:2003rj}, 
has remarkably boosted the multi-instanton calculus in gauge theories 
far beyond the results obtained in the past 
with standard methods, and many non-perturbative phenomena
can now be put in a framework amenable of a proper mathematical treatment. 
Recently, these localization techniques have been 
successfully applied also to multi-instantons of exotic type yielding results that are in perfect agreement with those expected from the heterotic/Type I$^\prime$ duality \cite{Billo:2009di}-\nocite{Fucito:2009rs}\cite{Billo:2010bd}
or from F-theory considerations \cite{Billo:2010mg}. It is therefore fair to say that
also the stringy multi-instanton calculus is now on a rather solid ground. 

In all examples of exotic multi-instantons considered up to now,
the gauge theory is realized either on the world-volume of D7 branes \cite{Billo:2009di,Fucito:2009rs}
or on systems of D7 and D3 branes \cite{Billo:2010bd,Billo:2010mg}; therefore, 
part of the results that have been obtained so far have necessarily an eight-dimensional 
interpretation due to the presence of the D7 branes. 
In this paper, instead, we consider a gauge sector made entirely of D3 branes so that
the results we get have only a four-dimensional character. In particular, we investigate the
gauge theory engineered with stacks of
fractional D3 branes in a $\mathbb{C}^3/\mathbb{Z}_3$ orientifold of type IIB 
preserving $\cN=2$ supersymmetry in four dimensions, and study 
the corresponding stringy multi-instanton configurations along the lines already discussed in \cite{Argurio:2007vqa} for the 1-instanton case. More specifically, 
we analyze a configuration of fractional
D3 branes that realizes an $\cN=2$ $\mathrm{SU}(N)$ theory in four dimensions with a hypermultiplet
in the symmetric representation, and then introduce exotic instantons 
by adding stacks of D(--1) branes on the nodes of the $\mathbb{C}^3/\mathbb{Z}_3$ quiver diagram
that are not occupied by the D3 branes. In this way the mixed open strings stretched
between the D3 and the D(--1) branes have only fermionic charged zero-modes, a typical feature of the
exotic instantons. Furthermore,
the orientifold projection removes the dangerous neutral fermionic zero-modes we alluded to above,
so that the stringy instantons have the right content of zero-modes to provide non-vanishing
contributions to the D3 brane effective action. 
We have computed such non-perturbative effects with the same localization methods \cite{Nekrasov:2002qd}-\nocite{Losev:2003py}\cite{Nekrasov:2003rj}
used to find the gauge instanton terms in the $\cN=2$ super Yang-Mills theory predicted by the Seiberg-Witten curve \cite{Seiberg:1994rs,Seiberg:1994aj}. However, due to the different
structure of the moduli space of the stringy instantons and of the corresponding moduli integrals,
the non-perturbative terms we obtain are of a novel type.

This paper is organized as follows. In Section \ref{sec:D3} we review the main features of the fractional D3 branes in the $\mathbb{C}^3/\mathbb{Z}_3$ orientifold and of the $\cN=2$ gauge theory
living on their world-volume. 
In Section \ref{sec:dinst} we introduce unoriented fractional D-instantons, focusing then
in Section \ref{sec:stringy} on the exotic configurations, on their moduli
spectrum and on the cohomological properties of their moduli action. In Section \ref{sec:stringyprepot} 
we explicitly evaluate the moduli integrals for the SU(2) theory,
and derive the non-perturbative corrections to the effective prepotential 
induced by the stringy instantons up to instanton number
$k=5$. Finally, in Section \ref{sec:concl} we summarize our results and present our conclusions.
Several technical details that are useful to reproduce some
of the computations of the main text are collected in the Appendix.

\section{D3 branes in the $\mathbb{C}^3/\mathbb{Z}_3$ orientifold}
\label{sec:D3}

We consider fractional D3 branes in a $\mathbb{C}^3/\mathbb{Z}_3$ orientifold and 
study the non-perturbative effects produced by fractional
D-instantons along the lines discussed in \cite{Argurio:2007vqa}. 
Even though this is quite standard, we briefly recall the main features 
of this orientifold model in order to be self-contained.

We place both the D3's and the D(--1)'s at the orbifold 
singularity, and parametrize the world-volume directions of the D3's 
by the first four string coordinates, as shown in Tab.~1.
\begin{table}[ht]
\begin{center}
\begin{tabular}{c|cccc|cccccc}
\phantom{\vdots}
&0&1&2&3&4&5&6&7&8&9 
\\
\hline
\phantom{\vdots}D3&$-$&$-$&$-$&$-$&$\times$&$\times$&$\times$&$\times$&$\times$&$\times$\\
\phantom{\vdots}D(--1)&$\times$&$\times$&$\times$&$\times$
&$\times$&$\times$&$\times$&$\times$&$\times$&$\times$\\
\end{tabular}
\end{center}
\label{tab:d3d-1}
\caption{D brane arrangement. The symbols $-$ and $\times$ denote respectively 
Neumann and Dirichlet boundary conditions for the open strings attached to the branes.}
\end{table} 
In the six-dimensional ``internal'' space orthogonal to the D3 branes 
we introduce three complex coordinates
\begin{equation}
 z^1=x^4+\ii x^5~,~~~~ z^2=x^6+\ii x^7~,~~~~ z^3=x^8+\ii x^9~,
\label{zs}
\end{equation}
on which the $\mathbb Z_3$ orbifold action can be naturally defined. Denoting by $g$ the
generator of $\mathbb Z_3$ such that $g^3=1$, we take
\begin{equation}
 \label{gorb}
g~:~~\begin{pmatrix}z^1 \cr  z^2 \cr z^3 \end{pmatrix}
~\to~
\begin{pmatrix}\xi\,z^1 \cr  \xi^{-1}\,z^2 \cr z^3 \end{pmatrix}
\end{equation}
where $\xi=\ee^{\frac{2\pi\ii}{3}}$.
Since one of the complex coordinates does not transform, this $\mathbb Z_3$ action breaks
half of the supersymmetries of the original ten-dimensional background and therefore leads
to $\mathcal N=2$ theories on the world-volume of the fractional D3 branes.

Notice that the action (\ref{gorb}) can be interpreted as a rotation of $+\frac{2\pi\ii}{3}$
in the $z^1$-plane combined with a rotation of $-\frac{2\pi\ii}{3}$ in the $z^2$-plane. Thus,
$g$ can be represented by
\begin{equation}
 \label{gorb1}
R(g) = \ee^{+\frac{2\pi\ii}{3}J_1}\,\ee^{-\frac{2\pi\ii}{3}J_2}
\end{equation}
where $J_i$ is the generator of the rotations in the $z^i$-plane in the vector representation. 
This expression is particularly useful to define the orbifold action
on spin-fields and, more generally, on fields carrying spinor indices. To this aim, in fact, it is
enough to take (\ref{gorb1}) with the generators $J_i$ in the spinor representation. 
As a consequence of the 
$4+6$ splitting of the ten-dimensional space-time induced by the D3 branes, the ``Lorentz'' group%
\footnote{Since we will be interested in studying instanton corrections, we take a Euclidean
signature in space-time.}
SO(10) is broken to $\mathrm{SO}(4)\times\mathrm{SO}(6)$, and thus any ten-dimensional
spinor decomposes accordingly. For example an anti-chiral spinor $\Lambda$ decomposes
as
\begin{equation}
 \Big(\Lambda^{\alpha A}\,,\,\Lambda_{\dot{\alpha} A}\Big)
\label{decospin}
\end{equation}
where $\alpha$ ($\dot\alpha$) are chiral (anti-chiral) spinor indices of SO(4), 
and the lower (upper) indices $A$ are chiral (anti-chiral) spinor indices of SO(6).
Upon using the explicit expression for the SO(6) spinor weights, from (\ref{gorb1}) we can
easily deduce that
\begin{equation}
 \label{gorbspin}
g~:~~\begin{pmatrix}\Lambda^{\alpha ---} \cr  \Lambda^{\alpha ++-} \cr \Lambda^{\alpha +-+}
\cr \Lambda^{\alpha -++} \end{pmatrix}
~\to~
\begin{pmatrix}\Lambda^{\alpha ---}\cr  \Lambda^{\alpha ++-} \cr \xi\,\Lambda^{\alpha +-+}
\cr \xi^{-1}\,\Lambda^{\alpha -++} \end{pmatrix}~~~\mbox{and}~~~
\begin{pmatrix}\Lambda_{\dot\alpha +++} \cr  \Lambda_{\dot\alpha --+} \cr \Lambda_{\dot\alpha -+-}
\cr \Lambda_{\dot\alpha +--} \end{pmatrix}
~\to~
\begin{pmatrix}\Lambda_{\dot\alpha +++}\cr  \Lambda_{\dot\alpha --+} \cr \xi^{-1}
\,\Lambda_{\dot\alpha -+-}
\cr \xi\,\Lambda_{\dot\alpha +--} \end{pmatrix}~.
\end{equation}
This action shows that only half of the spinor components are invariant under the orbifold action, 
thus leading to $\mathcal N=2$ supersymmetry as anticipated above.

The orbifold group $\mathbb Z_3$ has three irreducible representations: $R_1(g)=1$, $R_2(g)=\xi$ and $R_3(g)=\xi^{-1}$. Consequently \cite{Douglas:1996sw}, there are three
types of fractional D branes and the associated quiver diagram has three nodes. 
The number of fractional D3 branes occupying the $i$-th node which
corresponds to the representation $R_i(g)$ is denoted by $N_i$.
A generic open string excitation in this brane system
carries a Chan-Paton (CP) factor $X$
that is a $(N_1+N_2+N_3)\times(N_1+N_2+N_3)$ matrix on which the orbifold generator
$g$ acts according to
\begin{equation}
 \label{gCP}
g~:~~X~\to~\gamma(g)\,X\,\gamma(g)^{-1}~.
\end{equation}
Here $\gamma(g)$ is 
\begin{equation}
 \gamma(g) = 
\begin{pmatrix}
\one_{N_1}&0&0\cr  
0&\xi\,\one_{N_2}&0\cr
0&0&\xi^{-1}\,\one_{N_3}
\end{pmatrix}
\end{equation}
with $\one_{N_i}$ denoting the $N_i\times N_i$ identity matrix. 
This system supports an $\mathcal N=2$ gauge theory with group 
$\mathrm{U}(N_1)\times \mathrm{U}(N_2) \times \mathrm{U}(N_3)$ represented by the
quiver diagram of Fig.~1.

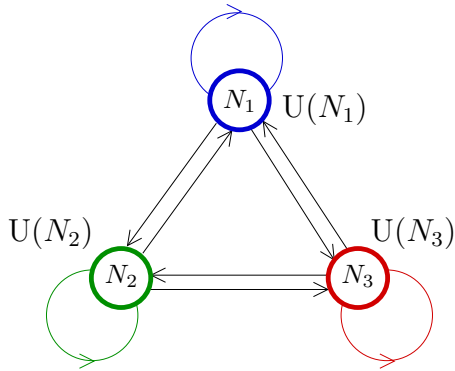
\begin{figure}[hbt]
  \begin{center}
\vspace{10pt}
\hspace{-30pt}
\begin{picture}(0,0)%
\includegraphics{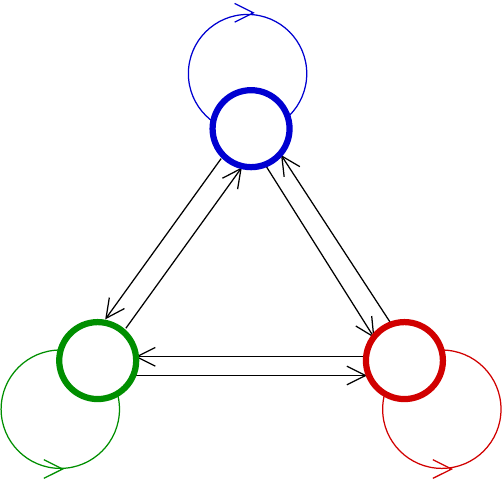}%
\end{picture}%
\setlength{\unitlength}{1450sp}%
\begingroup\makeatletter\ifx\SetFigFontNFSS\undefined%
\gdef\SetFigFont#1#2#3#4#5{%
  \reset@font\fontsize{#1}{#2pt}%
  \fontfamily{#3}\fontseries{#4}\fontshape{#5}%
  \selectfont}%
\fi\endgroup%
\begin{picture}(6030,5737)(541,-4460)
\put(1351,-2281){\makebox(0,0)[rb]{\smash{{\SetFigFont{12}{9.6}
 {\familydefault}{\mddefault}{\updefault}$\mathrm{U}(N_2)$}}}}
\put(2111,-2990){\makebox(0,0)[rb]{\smash{{\SetFigFont{10}{9.6}
 {\familydefault}{\mddefault}{\updefault}$N_2$}}}}
\put(5991,-171){\makebox(0,0)[rb]{\smash{{\SetFigFont{12}{9.6}
 {\familydefault}{\mddefault}{\updefault}$\mathrm{U}(N_1)$}}}}
\put(7491,-2281){\makebox(0,0)[rb]{\smash{{\SetFigFont{12}{9.6}
 {\familydefault}{\mddefault}{\updefault}$\mathrm{U}(N_3)$}}}}
\put(6111,-2990){\makebox(0,0)[rb]{\smash{{\SetFigFont{10}{9.6}
 {\familydefault}{\mddefault}{\updefault}$N_3$}}}}
\put(4111,-10){\makebox(0,0)[rb]{\smash{{\SetFigFont{10}{9.6}
 {\familydefault}{\mddefault}{\updefault}$N_1$}}}}
\end{picture}%
  \end{center}
 \label{Fig:1}
\caption{The $\mathbb{C}^3/\mathbb{Z}_3$ un-orientifolded theory corresponding to a configuration
of $N_1$, $N_2$ and $N_3$ fractional D3 branes. The lines starting and ending on the
same node represent $\cN=2$ vector multiplets in the adjoint representation of the $\mathrm{U}(N_i)$
groups. The oriented lines between different nodes represent bi-fundamental chiral multiplets
which pair up into $\cN=2$ hypermultiplets.}
 \end{figure}

We now enrich our configuration by adding an O3 plane with a world-volume lying along
the same four space-time directions as the D3 branes. The action of the orientifold
generator $\Omega$ on the various open string fields is standard and can be deduced by writing
\begin{equation}
 \Omega = \omega \,(-1)^{F_L}\,{\mathcal I}_{456789}
\label{OrientOperator}
\end{equation}
where $\omega$ is the world-sheet parity, $F_L$ the (left) space-time fermion number 
and ${\mathcal I}_{456789}$ is the reflection in the
internal space. On the other hand, the orientifold acts on the CP factors $X$ by means of a
matrix $\gamma(\Omega)$ according to 
\begin{equation}
 \label{omegaCP}
\Omega~:~~X~\to~\gamma(\Omega)\,X^{T}\,\gamma(\Omega)^{-1}~.
\end{equation}
In the presence of an orbifold the matrix $\gamma(\Omega)$ must satisfy the following
consistency condition \cite{Gimon:1996rq,Douglas:1996sw}
\begin{equation}
 \gamma(h)\, \gamma(\Omega) \,\gamma(h)^T = \gamma(\Omega)
\label{consistency}
\end{equation}
for any $h$ belonging to the orbifold group, which amounts to requiring that the orientifold
and orbifold projections commute with each other. 
The matrix $\gamma(\Omega)$ can be either symmetric or
antisymmetric. Here we choose to perform an antisymmetric projection on the D3 branes and denote
the corresponding matrix by $\gamma_-(\Omega)$. Taking $N_1$ to be even and $N_2=N_3$, we can
write
\begin{equation}
 \label{gamma-}
\gamma_-(\Omega) = 
\begin{pmatrix}
\epsilon&0&0\cr  
0&0&\one_{N_2}\cr
0&-\one_{N_2}&0
\end{pmatrix}
\end{equation}
where $\epsilon$ is a $N_1\times N_1$ antisymmetric matrix obeying $\epsilon^2=-1$. Using (\ref{gCP})
it is easy to verify that $\gamma_-(\Omega)$ satisfies the consistency condition 
(\ref{consistency}).

The bosonic field content on the fractional D3 branes at the singularity follows after
implementing the following orbifold and orientifold conditions%
\footnote{This same analysis can be performed in a straightforward way also in the fermionic sectors.}
\begin{subequations}
\begin{align}
 \mathbf{A}_\mu &= \gamma(g)\, \mathbf{A}_\mu \,\gamma(g)^{-1}~,~~~\phantom{\!(\xi)^I}
\mathbf{A}_\mu =- \gamma_-(\Omega) \,\big(\mathbf{A}_\mu\big)^T \,\gamma_-(\Omega)^{-1}~,
\label{vector_cond} \\
\mathbf{\Phi}^I &= (\xi)^I\,\gamma(g)\, \mathbf{\Phi}^I \,\gamma(g)^{-1}~,~~~
\mathbf{\Phi}^I =- \gamma_-(\Omega) \,\big(\mathbf{\Phi}^I\big)^T \,\gamma_-(\Omega)^{-1}~.
\label{scal_cond}
\end{align}
\end{subequations}
Here $\mathbf{A}_\mu$  is the gauge vector field along the D3 world-volume
directions ($\mu=0,\ldots,3$), while $\mathbf{\Phi}^I$ ($I=1,2,3$) are three complex scalars 
along the three complex directions (\ref{zs}).
The orbifold part of these conditions forces $\mathbf{A}_\mu$ and
$\mathbf{\Phi}^3$ to be $3\times 3$ block diagonal matrices, namely
\begin{equation}
 \mathbf{A}_\mu = \left(\begin{array}{ccc}
               A_{\mu(11)} & 0 & 0\\
               0 & A_{\mu(22)} & 0\\
               0 & 0 & A_{\mu(33)}
              \end{array}\right) ~,\ \ \ \ \ \ \ \ \ 
\mathbf{\Phi}^3 = \left(\begin{array}{ccc}
               \Phi^3_{(11)} & 0 & 0\\
               0 & \Phi^3_{(22)} & 0\\
               0 & 0 & \Phi^3_{(33)}
              \end{array}\right) ~,
\label{Amuphi3}
\end{equation}
and $\mathbf{\Phi}^1$ and $\mathbf{\Phi}^2$ to have the following off-diagonal structure
\begin{equation}
 \mathbf{\Phi}^1 = \left(\begin{array}{ccc}
               0& \Phi^1_{(12)}& 0\\
               0 & 0& \Phi^1_{(23)}\\
               \Phi^1_{(31)} & 0 & 0
              \end{array}\right) ~,\ \ \ \ \ \ \ \ \ 
\mathbf{\Phi}^2 = \left(\begin{array}{ccc}
               0& 0 & \Phi^2_{(13)}\\
               \Phi^2_{(21)} & 0 & 0\\
               0 & \Phi^2_{(32)} & 0
              \end{array}\right) ~.
\label{Phi1Phi2}
\end{equation}
The orientifold conditions impose that $A_{\mu(11)} = \epsilon \big(A_{\mu(11)}\big)^T
\!\epsilon$ and $A_{\mu(22)}=-\big(A_{\mu(33)}\big)^T$, and similarly that 
 $\Phi^3_{(11)} = \epsilon \big(\Phi^3_{(11)}\big)^T
\!\epsilon$ and $\Phi^3_{(22)}=-\big(\Phi^3_{(33)}\big)^T$. The resulting theory is therefore 
an $\mathrm{USp}(N_1)\times \mathrm{U}(N_2)$ gauge theory, with $\mathbf{A}_\mu$
and $\mathbf{\Phi}^3$ being the bosonic components 
of the ${\mathcal N}=2$ adjoint vector multiplet. 
Sometimes, it is convenient to still denote diagrammatically 
$A_{\mu(22)}$ and $A_{\mu(33)}$ (as well as 
$\Phi^3_{(22)}$ and $\Phi^3_{(33)}$) as belonging to different quiver nodes,
with the understanding that they should be actually identified in the above way.
Most often we will use the simplified notation $A_{\mu(22)}\equiv A_{\mu}$ and
$\Phi^3_{(22)}\equiv\Phi$.

The orientifold projection on the complex fields $\mathbf{\Phi}^1$ and $\mathbf{\Phi}^2$,
which represent the bosonic components of the matter superfields, can be done in a 
similar way and leads to the following relations
\begin{equation}
\Phi^1_{(12)} = -\epsilon \,\big(\Phi^1_{(31)}\big)^T~,~~ 
\Phi^1_{(23)} = \big(\Phi^1_{(23)}\big)^T~,~~ 
\Phi^2_{(13)} = \epsilon \,\big(\Phi^2_{(21)}\big)^T,~~
\Phi^2_{(32)} = \big(\Phi^2_{(32)}\big)^T~.
\label{phi12}
\end{equation}
With respect to the gauge group $\mathrm{USp}(N_1)\times\mathrm{U}(N_2)$ they belong to the representations given in Tab.~2.
\begin{table}[ht]
\begin{center}
\begin{tabular}{|c||c|c|}
\hline
\phantom{\vdots}
field &$\mathrm{USp}(N_1)$&$\mathrm{U}(N_2)$ 
\\
\hline\hline
$\phantom{\vdots}\Phi^1_{(12)}$ & $\Yfund$ &$\overline{\Yfund}$
\\
$\phantom{\vdots}\Phi^1_{(31)}$ & $\Yfund$ &${\Yfund}$
\\
$\phantom{\vdots}\Phi^1_{(23)}$ & \begin{LARGE}${\cdot}$\end{LARGE} &${\Ysymm}$
\\
$\phantom{\vdots}\Phi^2_{(21)}$ & $\Yfund$ &${\Yfund}$
\\
$\phantom{\vdots}\Phi^2_{(13)}$ & $\Yfund$ &$\overline{\Yfund}$
\\
$\phantom{\vdots}\Phi^2_{(32)}$ & \begin{LARGE}${\cdot}$\end{LARGE} &$\overline{\Ysymm}$
\\
\hline
\end{tabular}
\end{center}
\label{tab:matter}
\caption{Matter content and associated gauge representations.}
\end{table} 

In the following we will consider a D3 brane system with $N_1=0$ and $N_2=N_3=N$, supporting a
four-dimensional gauge theory with group $\mathrm{U}(N)$ 
and matter in the symmetric representation.
Actually, we can neglect the U(1) factor since it is IR free, and thus we will concentrate
only on the low-energy dynamics of the $\mathrm{SU}(N)$ part. Note that the complex
fields $\Phi^1_{(23)}$ and $\overline{\Phi}^{\,2}_{(32)}$, plus their fermionic partners,
pair up and build an $\mathcal N=2$ hypermultiplet in the symmetric representation of
$\mathrm{SU}(N)$. For such a gauge theory, the 1-loop $\beta$-function coefficient is
\begin{equation}
 \label{beta}
b_1=N-2~.
\end{equation}
The theory is therefore UV asymptotically free for $N>2$ and conformal for $N=2$. The latter case
is a non-standard realization of the $\mathcal N=4$ SU(2) superconformal 
Yang-Mills theory; indeed for SU(2)
the symmetric representation coincides with the adjoint, and thus the matter hypermultiplet
can be combined with the vector multiplet enhancing the supersymmetry from $\mathcal N=2$
to $\mathcal N=4$. In the following we will see that this realization leads to non-trivial
results in the non-perturbative sectors of the theory even in the superconformal case.

\section{D-instantons in the $\mathbb{C}^3/\mathbb{Z}_3$ orientifold}
\label{sec:dinst}

We now briefly discuss the D-instantons in the $\mathbb{C}^3/\mathbb{Z}_3$ orientifold
introduced in the previous section. The most general instanton configuration is realized by
putting $k_1$ D(--1) branes on node 1, $k_2$ D(--1)'s on node 2 and $k_3$ D(--1)'s on node
3 with $k_2=k_3$. A generic open string excitation stretching between two D-instantons
will therefore have a CP factor $Y$ which is a $(k_1+2k_2)\times(k_1+2k_2)$ matrix. On it the
$\mathbb Z_3$ orbifold generator $g$ acts by means of a matrix $\gamma'(g)$ which has the same 
form as $\gamma(g)$ in 
(\ref{gCP}) but with $N_i$ replaced with $k_i$. 
The orientifold action on the D(--1) CP factors is instead different with respect to the
D3 case \cite{Gimon:1996rq,Argurio:2007vqa}. 
Indeed, the consistency with the antisymmetric matrix 
(\ref{gamma-}) chosen for the D3 branes
requires to transform the D(--1) CP factors with a symmetric matrix
$\gamma_+(\Omega)$ according to
\begin{equation}
 \label{omegaCPinst}
\Omega~:~~Y~\to~\gamma_+(\Omega)\,Y^{T}\,\gamma_+(\Omega)^{-1}
\end{equation}
where%
\footnote{Notice that, differently from $N_1$, $k_1$ does not need to be even.}
\begin{equation}
 \label{gamma+}
\gamma_+(\Omega) = 
\begin{pmatrix}
\one_{k_1}&0&0\cr  
0&0&\one_{k_2}\cr
0&\one_{k_2}&0
\end{pmatrix}~.
\end{equation}
Adopting an ADHM-inspired notation, we can organize the bosonic excitations 
in the Neveu-Schwarz sector of the open strings suspended between two D-instantons in a four-dimensional 
vector $\mathbf{a}_\mu$ and three complex scalars $\bm{\chi}^I$,
which are subject to the following conditions
\begin{subequations}
\label{orien-1}
\begin{align}
 \mathbf{a}_\mu &= \gamma'(g)\, \mathbf{a}_\mu \,\gamma'(g)^{-1}~,~~~~\phantom{\!(\xi)^I}
\mathbf{a}_\mu =+ \gamma_+(\Omega) \,\big(\mathbf{a}_\mu\big)^T \,\gamma_+(\Omega)^{-1}~,
\label{amu} \\
\bm{\chi}^I &= (\xi)^I\,\gamma'(g)\, \bm{\chi}^I \,\gamma'(g)^{-1}~,~~~
\bm{\chi}^I =- \gamma_+(\Omega) \,\big(\bm{\chi}^I\big)^T \,\gamma_+(\Omega)^{-1}~.
\label{chiI}
\end{align}
\end{subequations}
The plus sign in the orientifold action on $\mathbf{a}_\mu$ is due to the fact that now
the first four directions labeled by $\mu$ are of Dirichlet type. Implementing the
constraints (\ref{orien-1}) we obtain
\begin{equation}
 \mathbf{a}_\mu = \left(\begin{array}{ccc}
               a_{\mu(11)} & 0 & 0\\
               0 & a_{\mu(22)} & 0\\
               0 & 0 & a_{\mu(33)}
              \end{array}\right) ~,\ \ \ \ \ \ \ \ \ 
\bm{\chi}^3 = \left(\begin{array}{ccc}
               \chi^3_{(11)} & 0 & 0\\
               0 & \chi^3_{(22)} & 0\\
               0 & 0 & \chi^3_{(33)}
              \end{array}\right) ~,
\label{amuchi}
\end{equation}
with
\begin{equation}
 a_{\mu(11)} = \big(a_{\mu(11)}\big)^T~,~~a_{\mu(22)}=\big(a_{\mu(33)}\big)^T~,~~
 \chi^3_{(11)} = -\big(\chi^3_{(11)}\big)^T~,~~\chi^3_{(22)}=-\big(\chi^3_{(33)}\big)^T~,
\label{amuchi1}
\end{equation}
and
\begin{equation}
 \bm{\chi}^1 = \left(\begin{array}{ccc}
               0& \chi^1_{(12)}& 0\\
               0 & 0& \chi^1_{(23)}\\
               \chi^1_{(31)} & 0 & 0
              \end{array}\right) ~,\ \ \ \ \ \ \ \ \ 
\bm{\chi}^2 = \left(\begin{array}{ccc}
               0& 0 & \chi^2_{(13)}\\
               \chi^2_{(21)} & 0 & 0\\
               0 & \chi^2_{(32)} & 0
              \end{array}\right) ~,
\label{chi1chi2}
\end{equation}
with
\begin{equation}
 \chi^1_{(12)} = -\big(\chi^1_{(31)}\big)^T~,~~\chi^1_{(23)}=-\big(\chi^1_{(23)}\big)^T~,~~
 \chi^2_{(13)} = -\big(\chi^2_{(21)}\big)^T~,~~\chi^2_{(32)}=-\big(\chi^2_{(32)}\big)^T~.
\label{chichi}
\end{equation}
The conditions (\ref{amuchi1}) imply that the symmetry group on the D-instantons
is $\mathrm{SO}(k_1)\times\mathrm{U}(k_2)$, with the orthogonal factor referring to the first node
of the quiver and the unitary factor to the remaining two nodes that are identified with each other
under the orientifold projection.

This analysis can be easily extended also to the fermionic excitations of the Ramond sector. We will provide some details on this in the following sections. Here, instead, we dwell on the fact that 
depending on whether or not the D-instanton occupies a quiver node 
populated also by a stack of D3 branes,
it represents an ordinary gauge instanton or a stringy instanton.
Referring to the $\mathrm{SU}(N)$ theory of the previous section, which corresponds to
a D3 brane configuration of type $(N_1,N_2)=(0,N)$, a D-instanton configuration of type
$(k_1,k_2)=(0,k)$ describes a gauge instanton with instanton number $k$ and instanton group
$\mathrm{U}(k)$. On the other hand, a D-instanton configuration of type $(k_1,k_2)=(k,0)$ describes a
stringy instanton with charge $k$ and instanton group $\mathrm{SO}(k)$.%
\footnote{The occurrence of an orthogonal
symmetry in the instanton sector of a theory with a unitary gauge group is a clear signal of the exotic character
of the stringy instantons.}
All this is summarized
in Tab.~3. 
\begin{table}[ht]
\begin{center}
\begin{tabular}{c|ccc|c|c}
\phantom{\vdots}
&D3's&$\oplus$&D(--1)'s&gauge group&instanton group
\\
\hline
\phantom{\vdots}gauge instantons~~~&$(0,N)$&$\oplus$&$(0,k)$&SU($N$)&U($k$)
\\
\phantom{\vdots}stringy instantons~~~&$(0,N)$&$\oplus$&$(k,0)$&SU($N$)&SO($k$)
\end{tabular}
\end{center}
\label{tab:inst}
\caption{D3 and D(--1) brane configurations and their associated symmetry groups corresponding to
gauge and exotic instantons.}
\end{table} 

The most general D-instanton configuration for our $\mathrm{SU}(N)$ gauge theory is therefore
a superposition of gauge and stringy instantons. In the following sections we will discuss in
detail the spectrum of moduli for the stringy instantons, and compute explicitly 
their contributions to the gauge effective action for $N=2$. The analysis for $N>2$ will
be presented in a separate publication \cite{progress}.

\section{Stringy instantons}
\label{sec:stringy}

We now describe in more detail the stringy instanton configurations and thus consider
a system made of a stack of $k$ D(--1) branes placed on node $1$ 
of the quiver diagram and two stacks of $N$ D3 branes placed on nodes $2$ and $3$ and identified 
with each other under the orientifold action. 

\subsection{Moduli spectrum}
The open strings excitations with at least one end-point on the D-instantons
can be distinguished into neutral and charged
ones, which we are going to analyze in turn.

\paragraph{Neutral sector:} The neutral sector contains the modes of the open strings starting and
ending on the D-instantons which are therefore uncharged under the gauge group of the D3 branes.
Since in this configuration there is only one stack of instantonic branes on node $1$, the
CP factors of the neutral moduli have only one non-zero entry, {\it i.e.} the $(11)$ component
which is a $k\times k$ matrix. Since the complex scalars $\bm{\chi^1}$ and $\bm{\chi^2}$
do not have a $(11)$ component as is clear from \eq{chi1chi2}, we can set 
$\bm{\chi^1} = \bm{\chi^2}=0$. Furthermore, for the moduli $\mathbf{a}_\mu$ and $\bm{\chi}^3$
which do have a diagonal $(11)$ component in their CP factors, 
we can simplify the notation and put
\begin{equation}
 a_{\mu(11)}\equiv a_\mu=\big(a_\mu\big)^T~,~~~~\chi^3_{(11)}\equiv \chi=-\big(\chi\big)^T~.
\end{equation}

As far as the fermionic moduli are concerned, we see from the spinor transformation
properties (\ref{gorbspin}) that only the components with 
indices $(\alpha---)$, $(\alpha++-)$, $(\dot\alpha+++)$ and $(\dot\alpha--+)$ 
are invariant under the $\mathbb Z_3$ orbifold action. Therefore, in the configuration 
we are now considering, these are the only components that can have a $(11)$ entry in their 
CP factors and can then survive the orbifold projection. Adopting an
ADHM inspired notation, we denote them as $M^{\alpha a}$ and $\lambda_{\dot\alpha a}$
where the upper index $a$ takes the values $(---)$ and $(++-)$, while the
lower index $a$ takes the values $(+++)$ and $(--+)$. Also these fermionic moduli 
are $k\times k$ matrices and on them the orientifold projection acts according to
\begin{equation}
M^{\alpha a} = +\big(M^{\alpha a}\big)^T 
~,~~~
\lambda_{\dot\alpha a} = -\big(\lambda_{\dot\alpha a} 
\big)^T ~.
\label{mlambda}
\end{equation}
These rules are a consequence of the fact that the matrix $\gamma_+(\Omega)$ restricted to the
$(11)$ block of a CP factor is simply the identity and that the orientifold 
generator (\ref{OrientOperator}) acting on a ten-dimensional spinor effectively 
measures its chirality in the first four directions, from which the signs in (\ref{mlambda}) 
immediately follow.

\paragraph{Charged sector:} The charged sector contains the modes of the open strings which
have one end-point on the D-instantons and one on the D3 branes, and which are charged
under the gauge group created by the latter. Since in the exotic configuration the
D-instantons sit on node 1 while the D3 branes occupy nodes 2 and 3, the
CP factors for the 3/(--1) strings and the (--1)/3 strings have, respectively, the
following structure
\begin{equation}
\begin{pmatrix}
\,0\,&\,0\,&\,0\,\cr
\star&\,0\,&\,0\,\cr
\star&\,0\,&\,0\,
\end{pmatrix} ~~~~\mbox{and}~~~~\begin{pmatrix}
\,0\,&\star&\star\cr
\,0\,&\,0\,&\,0\,\cr
\,0\,&\,0\,&\,0\,
\end{pmatrix}~.
\label{3-1}
\end{equation}
It is easy to realize that both such CP factors transform non-trivially under 
the orbifold generator $g$ represented by the matrices $\gamma(g)$ and $\gamma'(g)$.
Thus, the only charged states surviving the orbifold projection are
those whose vertex operators transform under $g$ in such a way to compensate
the phase acquired by their CP factors. In the Neveu-Schwarz sector, due to the
mixed Neumann-Dirichlet boundary conditions, the
GSO projected physical vertex operators carry an anti-chiral spinor index in the first four
directions but are singlets in the internal directions where the orbifold acts.
Thus, these bosonic vertex operators do not acquire any phase under $g$ and cannot
survive the orbifold projection for the above argument. The absence of bosonic
charged moduli is a typical signal of the exotic nature of these instanton configurations.
On the other hand, in the Ramond sector, the GSO projected
physical vertex operators are anti-chiral spinors in the six internal directions and
two of their components, namely those with indices $(+-+)$ and $(-++)$,
transform non-trivially under $g$ as one can see from (\ref{gorbspin}), and can survive the
orbifold projection.
Being more explicit and adopting again an ADHM inspired notation, 
the physical charged moduli of the 3/(--1) sector are
\begin{equation}
\bm{\mu}^{+-+}=\begin{pmatrix}
\,0\,&\,0\,&\,0\,\cr
\,0\,&\,0\,&\,0\,\cr
\mu&\,0\,&\,0\,
\end{pmatrix} ~~~~\mbox{and}~~~~
\bm{\mu}^{-++}=\begin{pmatrix}
\,0\,&\,0\,&\,0\,\cr
\mu'&\,0\,&\,0\,\cr
\,0\,&\,0\,&\,0\,
\end{pmatrix}
 \label{mumu'}
\end{equation}
where both $\mu$ and $\mu'$ are $N\times k$ matrices. The physical moduli in the
(--1)/3 sector, corresponding to open strings with opposite orientation, are related
to those of the 3/(--1) sector through the orientifold action. In our case we have
\begin{equation}
 \begin{aligned}
  \bm{\bar\mu}^{+-+}&= \gamma_+(\Omega)\big(\bm{\mu}^{+-+}\big)^T\gamma_-(\Omega)^{-1}
= \begin{pmatrix}
\,0\,&+\mu^T&\,0\,\cr
\,0\,&\,0\,&\,0\,\cr
\,0\,&\,0\,&\,0\,
\end{pmatrix}~,\\
\bm{\bar\mu}^{-++}&= \gamma_+(\Omega)\big(\bm{\mu}^{-++}\big)^T\gamma_-(\Omega)^{-1}
= \begin{pmatrix}
\,0\,&\,0\,&-\mu^{\prime T}\cr
\,0\,&\,0\,&\,0\,\cr
\,0\,&\,0\,&\,0\,
\end{pmatrix}~.
 \end{aligned}
\label{barmumu'}
\end{equation}

\subsection{Moduli action}
As shown in \cite{Green:2000ke,Billo:2002hm} 
the moduli action can be obtained from open string disk-amplitudes involving all moduli listed
above.
Such action can be expressed as the sum of three parts,
\begin{equation}
 S = S_{1} + S_{2} + S_{3}~,
\label{s}
\end{equation}
with
\begin{subequations}
 \begin{align}
S_{1} &=\frac{1}{g_0^2} \,\tr 
\Big\{\!\!-\frac{1}{4} \big[a^\mu,a^\nu\big]\,\big[a_\mu,a_\nu\big] 
- \big[a_\mu,\chi\big]\,\big[a^\mu,\overline{\chi}\big]
+ \frac{1}{2} \,\big[\overline{\chi},\,\chi\big]\, \big[\overline{\chi},\chi\big]\,  \Big\} ~,
\label{quartic}\\
S_{2} &=\frac{1}{g_0^2} \, \tr \Big\{ 2\, \lambda_{\dot{\alpha}a} 
\big[a^\mu,M_\beta^{\ a}\big]\, (\overline{\sigma}_\mu)^{\dot{\alpha}\beta} 
- \ii\, \lambda_{\dot{\alpha}a} \big[\chi,\lambda^{\dot{\alpha}a}\big] 
- 2 \ii\, M^{\alpha a} \big[\overline{\chi},M_{\alpha a}\big] \Big\} ~,
\label{cubic}\\
S_{3} &=\frac{1}{g_0^2} \, \tr \Big\{\!\!-\ii\, \mu^T\mu'\,\chi\Big\}
 \label{mixed}
 \end{align}
\label{ss}
\end{subequations}
\!\!corresponding, respectively, to quartic, cubic and mixed interactions.
Here the trace is over the $\mathrm{SO}(k)$ indices and $g_0$ is the coupling constant of the zero-dimensional Yang-Mills theory on the D$(-1)$ branes,
which is related to the string coupling constant $g_s$ through the relation
\begin{equation}
 g_0^2 = \frac{g_s}{4 \pi^3 \alpha'^2} ~.
\label{g0}
\end{equation}
All moduli appearing in this action have canonical scaling dimensions, namely the bosons have dimension
of (length)$^{-1}$ and the fermions dimension of (length)$^{-3/2}$. More standard ADHM-dimensions can
be obtained absorbing suitable powers of $g_0$, but we refrain from doing this.

The quartic interaction terms among the $a_\mu$'s can be disentangled by means 
of the three auxiliary fields $D_c$
($c=1,2,3$), so that $S_{1}$ can be rewritten in the following way
\begin{equation}
 S'_{1} 
=\frac{1}{g_0^2} \,\tr \Big\{\frac{1}{2} D_c D^c 
- \frac{1}{2} D_c \overline{\eta}^c_{\mu\nu}\, \big[a^\mu,a^\nu\big] 
- \big[a_\mu,\chi\big]\,\big[a^\mu,\overline{\chi}\big]   
+ \frac{1}{2} \big[\overline{\chi},\,\chi\big]\,\big[\overline{\chi},\,\chi\big] \Big\}
\label{quartic1}
\end{equation}
where $\overline{\eta}^c_{\mu\nu}$ are the anti-self dual 't Hooft symbols.
Indeed, eliminating the auxiliary fields through their algebraic equations
\begin{equation}
D^c = \frac{1}{2} \, \overline{\eta}^c_{\mu\nu} [a^\mu,a^\nu]~,
\label{eqD}
\end{equation}
one can see that $S'_{1}$ reduces to $S_{1}$.

Another useful rewriting concerns the cubic action (\ref{cubic}). It is obtained
by making suitable combinations among the components of the fermionic moduli 
that correspond to a ``topological twist'' in which the internal spinor index $a$
is identified with a space-time spinor index $\dot\beta$. 
More explicitly, this identification leads to
\begin{equation}
\begin{aligned}
& \lambda_{\dot\alpha a}
~\to~\lambda_{\dot\alpha\dot\beta}\equiv\frac{1}{2}\, \epsilon_{\dot{\alpha}\dot{\beta}} \,\eta 
+ \frac{\ii}{2}\, (\tau^c)_{\dot{\alpha}\dot{\beta}}\, \lambda_c ~,\\
& M^{\alpha a}~\to~ M^{\alpha\dot\beta} \equiv\frac{1}{2}\, M_\mu \,(\sigma^\mu)^{\alpha\dot\beta}~. 
\end{aligned}
\label{eta}
\end{equation}
In this way the original Lorentz group $\mathrm{SU}(2)_L\times\mathrm{SU}(2)_R$ gets replaced by
the ``twisted'' version $\mathrm{SU}(2)\times\mathrm{SU}(2)'$ where $\mathrm{SU}(2) = \mathrm{SU}(2)_L$
and $\mathrm{SU}(2)'=\diag\big(\mathrm{SU}(2)_R,\mathrm{SU}(2)_I\big)$ with $\mathrm{SU}(2)_I$
being the internal $R$-symmetry group of the $\cN=2$ theory.

With the definitions (\ref{eta}), the cubic action $S_{2}$ can be rewritten as follows
\begin{equation}
  S'_{2} =\frac{1}{g_0^2} \, \tr \Big\{\! \eta \, \big[a_\mu,M^\mu\big]  \!
+\lambda_c \,\big[a^\mu,M^\nu\big] \, \overline{\eta}^c_{\mu\nu} 
\!- \frac{\ii}{2} \,\eta\, \big[\chi,\eta \big] \!-  \frac{\ii}{2} 
\,\lambda_c \,\big[\chi, \lambda^c \big] \!- \ii M_\mu \big[\overline{\chi},M^\mu\big] \Big\} ~.
\label{cubic1}
\end{equation}
Finally, it is also convenient to replace the mixed action (\ref{mixed}) with
\begin{equation}
  S'_{3} =\frac{1}{g_0^2} \, \tr \Big\{\!\!-\ii\, \mu^T\mu'\,\chi+h^Th'\Big\}
\label{mixed1}
\end{equation}
where $h$ and $h'$ are charged auxiliary fields which do not interact with any other modulus.
Even if this replacing looks trivial, it is nevertheless useful for reasons that will become
clear in a moment. 

The total action 
\begin{equation}
 S' = S'_{1} + S'_{2} + S'_{3}
\label{s'}
\end{equation}
is invariant under the D-instanton group $\mathrm{SO}(k)$ and the D3 brane
gauge group $\mathrm{SU}(N)$. It is also invariant under the ``twisted'' Lorentz
group $\mathrm{SU}(2)\times\mathrm{SU}(2)'$ under which $a_\mu$ and $M_\mu$ transform in the
$(\mathbf{2},\mathbf{2})$, $\lambda_c$ and $D_c$ in the $(\mathbf{1},\mathbf{3})$, and
all the remaining moduli $\chi$, $\overline{\chi}$, $\eta$, $\mu$, $\mu'$, $h$ and $h'$ are
singlets. Furthermore, the action (\ref{s'}) is invariant under the following BRST-like
transformations
\begin{equation}
 \label{Q}
\begin{aligned}
& Q a^\mu  = M^\mu~,~~~ Q M^\mu = \ii\comm{\chi}{a^\mu}~, \\
& Q\lambda_c = D_c~,~~~ Q D_c  =  \ii\comm{\chi}{\lambda_c}~,\\
& Q\, \overline{\chi} = -\ii\eta~,~~~ Q\eta  = -\comm{\chi}{\overline{\chi}}~,~~~ Q \chi = 0~, \\
& Q \mu = h~,~~~ Q h = \ii\,\mu\,\chi~,\\
& Q \mu' = h'~,~~~ Q h' = \ii\,\mu'\,\chi~.
\end{aligned}
\end{equation}
The BRST charge $Q$ is the ``singlet'' component of the supercharges $Q_{\dot\alpha a}$ that arises
after the topological twist that identifies $a$ with $\dot\beta$, namely
\begin{equation}
 Q \equiv Q_{\dot\alpha\dot\beta}\,\epsilon^{\dot\alpha\dot\beta}~.
\label{defQ}
\end{equation}
Note that $Q$ is nilpotent up to an infinitesimal $\mathrm{SO}(k)$ transformation 
parametrized by $\chi$. Indeed, on any modulus we have
\begin{equation}
 Q^2 \, \bullet = T_{\mathrm{SO}(k)}(\chi) \, \bullet~,
\label{Q2}
\end{equation}
where $T_{\mathrm{SO}(k)}(\chi)$ denotes an infinitesimal $\mathrm{SO}(k)$ rotation
with parameter $\chi$ in the appropriate representation.
According to (\ref{Q}), all moduli except $\chi$ form BRST doublets of the type $(\Psi_0,\Psi_1)$
such that $Q\,\Psi_0=\Psi_1$ and whose properties are collected in Tab.~4.
\begin{table}[ht]
\begin{center}
\begin{tabular}{|c||c|c|c|}
\hline
\phantom{\vdots}
$(\Psi_0,\Psi_1)$&$\mathrm{SO}(k)$ &$\mathrm{SU}(N)$&$\mathrm{SU}(2)\times\mathrm{SU}(2)'$
\\
\hline\hline
$\phantom{\vdots}(a_\mu,M_\mu)$ & $\Ysymm$ &$\mathbf{1}$&$(\mathbf{2},\mathbf{2})$
\\
$\phantom{\vdots}(\lambda_c,D_c)$ & $\Yasymm$
 &$\mathbf{1}$&$(\mathbf{1},\mathbf{3})$
\\
$\phantom{\vdots}(\overline{\chi},\eta)$ &$\Yasymm$ & $\mathbf{1}$
 &$(\mathbf{1},\mathbf{1})$
\\
$\phantom{\Big|}(\mu,h)$ & $\Yfund$ &$\mathbf{\overline{N}}$&$(\mathbf{1},\mathbf{1})$
\\
$\phantom{\Big|}(\mu',h')$ & $\Yfund$ &$\mathbf{\overline{N}}$&$(\mathbf{1},\mathbf{1})$
\\
\hline
\end{tabular}
\end{center}
\label{tab:rep}
\caption{Moduli in the stringy instanton configuration organized as BRST pairs and 
their transformation properties under the various symmetry groups.}
\end{table} 

By exploiting the above properties and using the invariance under $\mathrm{SO}(k)$, one
can easily show that the total action (\ref{s'}) is $Q$-exact; indeed
\begin{equation}
\label{S'1}
 S' = Q \,\Xi ~,
\end{equation}
with
\begin{equation}
 \Xi=\frac{1}{g_0^2} \, \tr \Big\{\ii M^\mu \big[\overline{\chi},a_\mu\big]
- \frac{1}{2}\, \overline{\eta}^c_{\mu\nu} \lambda_c \big[a^\mu,a^\nu\big]
+ \frac{1}{2} \,\lambda_c D^c 
- \frac{1}{2} \,\big[\chi,\overline{\chi}\big] \eta 
+ \mu^Th'\Big\}~.
\label{Xi}
\end{equation}

Since the scaling dimension of the BRST charge is (length)$^{-1/2}$, the dimensions of the
components $(\Psi_0,\Psi_1)$ of any BRST doublet are of the form (length)$^\Delta$ and (length)$^{\Delta-1/2}$. Thus, recalling
that a fermionic variable and its differential have opposite dimensions, 
the measure on the instanton moduli space 
\begin{equation}
 d\mathcal{M}_{k} \equiv \,d\chi \prod_{(\Psi_0,\Psi_1)} d\Psi_0\,d\Psi_1
\label{measure}
\end{equation}
has the total dimension
\begin{equation}
 \mbox{(length)}^{-\frac{1}{2}k(k-1)+\frac{1}{2}n_b-\frac{1}{2}n_f}~.
\label{dimmeas}
\end{equation}
Here, the first term in the exponent accounts for the unpaired modulus $\chi$ in the
anti-symmetric representation of $\mathrm{SO}(k)$, while $n_b$ ($n_f$) denotes the number
of BRST multiplets whose lowest components $\Psi_0$ are bosonic (fermionic). From
Tab.~4 it is not difficult to verify that $n_b=\frac{5}{2}\,k^2+\frac{3}{2}\,k$
and $n_f=\frac{3}{2}\,k^2-\frac{3}{2}\,k+2kN$, so that the measure (\ref{measure}) has dimension
\begin{equation}
 \mbox{(length)}^{k(2-N)}=\mbox{(length)}^{-kb_1}
\label{dimmeas1}
\end{equation}
where $b_1$ is the coefficient of the 1-loop $\beta$-function for our gauge theory, given 
in (\ref{beta}). The negative sign in the exponent of (\ref{dimmeas1}) is another hallmark
of the intrinsically stringy nature of the instanton configuration we are considering%
\footnote{For the usual gauge theory instantons the dimension of the moduli measure is $\mbox{(length)}^{+kb_1}$, see for instance \cite{Dorey:2002ik} for a general discussion.}.
However, in the conformal $N=2$ case which we will discuss in detail in the following section
also the exotic instanton measure (\ref{measure}) is dimensionless and thus one expects that
some non-perturbative contributions may be seen also in the effective field theory. In Sect. 
\ref{sec:stringyprepot} we will explicitly see that this is indeed what happens.

\subsection{Deformed moduli action}
To obtain the non-perturbative contributions induced by the stringy instantons, 
it is necessary to generalize the moduli action 
(\ref{S'1}) and fully exploit all
symmetries of the instanton moduli space, which are the gauge group $\mathrm{SO}(k)$
on the $k$ D(--1)'s, the gauge group $\mathrm{SU(N)}$ on the $N$ D3 branes and
the ``twisted'' Lorentz group $\mathrm{SU}(2)\times \mathrm{SU}(2)'$. 

To this aim, we begin by considering the interactions among the instanton moduli and 
the gauge fields propagating on the world-volume of the D3 branes, which we combine 
into an $\mathcal N=2$ chiral superfield ${\Phi}(x,\theta)$.
Such interactions can be easily obtained by computing mixed disk amplitudes
involving both vertex operators for moduli and vertex operators for dynamical fields, as
discussed in detail in \cite{Billo:2002hm,Billo:2006jm} 
for analogous D(--1)/D3 brane systems. In the present case the result of such computations
is 
\begin{equation}
 \label{intPhi}
\frac{1}{g_0^2}\, \tr \Big\{\ii\,\mu^T{\Phi}(x,\theta)\,\mu'\Big\}
\end{equation}
which has to be added to the moduli action (\ref{S'1}). For our later purposes it is enough to 
focus on the dependence on the vacuum expectation value
\begin{equation}
 \label{vev}
\phi = \langle {\Phi}(x,\theta) \rangle ~,
\end{equation}
and hence we will consider the following modified mixed action
\begin{equation}
 {S}'_{3}(\phi) = {S}'_{3} 
+ \frac{1}{g_0^2}\, \tr \Big\{\ii\,\mu^T\phi\,\mu'\Big\}~.
\label{smixed2}
\end{equation}

Another kind of deformation concerns the inclusion of a non-trivial background
to fully exploit the Euclidean Lorentz symmetry in the four space-time directions.
This is usually called the $\Omega$-background deformation \cite{Nekrasov:2002qd}-\nocite{Losev:2003py}\cite{Nekrasov:2003rj}
which, in our stingy context, can be realized by turning on a non-trivial 
Ramond-Ramond 3-form flux as discussed in detail
in \cite{Billo:2006jm} and more recently in 
\cite{Ito:2010vx} where the equivalence between the $\Omega$-background and the Ramond-Ramond flux has been shown in full generality. More specifically, we introduce a Ramond-Ramond 3-form
flux of the type $F_{\mu\nu z^3}$,
{\it i.e.} with two indices along the 4-dimensional world-volume of the D3 branes and 
one holomorphic index in the internal direction left invariant by the $\mathbb Z_3$ orbifold. 
It is not difficult to realize that such a field strength survives the orientifold projection
under $\omega\,(-1)^{F_L}\,\mathcal I_{456789}$, since
$F_{\mu\nu z^3}$ is even under the world-sheet parity $\omega$ 
(like any other RR 3-form field strength), 
odd under $(-1)^{F_L}$ (like any field of the Ramond-Ramond sector)
and odd under the inversion $\mathcal I_{456789}$ (like any field with only one index
in the internal directions). {From} now on, we denote $F_{\mu\nu z^3}$ simply as
$\mathcal F_{\mu\nu}$ and parametrize it in terms of the 't Hooft symbols
as follows
\begin{equation}
 \mathcal F_{\mu\nu} = -\frac{\ii}{2}\,\bar f_c\,\eta^c_{\mu\nu} -\frac{\ii}{2}\, 
f_c\,\overline{\eta}^c_{\mu\nu}~,
\label{fc}
\end{equation}
with $\bar f_c$ and $f_c$ belonging, respectively, to the representations $(\mathbf{3},\mathbf{1})$ 
and $(\mathbf{1},\mathbf{3})$ of $\mathrm{SU}(2)\times\mathrm{SU}(2)'$.
Furthermore, for reasons that will become apparent in the
following, we also turn on the component of the
Ramond-Ramond 3-form field-strength with an anti-holomorphic index in the internal space,
{\it i.e.} $F_{\mu\nu\bar{z}^3} \equiv \overline{\mathcal F}_{\mu\nu}$.
We then compute mixed disk amplitudes with insertions of $\mathcal F_{\mu\nu}$
and $\overline{\mathcal F}_{\mu\nu}$ to obtain their couplings with the instanton moduli. 
The results of these calculations, which are performed as explained in detail in
\cite{Billo:2006jm,Billo':2008sp,Billo:2009di}, are new terms in the moduli action that
can be accounted by replacing the quartic and cubic terms 
given in (\ref{quartic1}) and (\ref{cubic1}) as follows
\begin{equation}
 \begin{aligned}
S'_{1} & \rightarrow S'_{1}(\mathcal F,\overline{\mathcal F}) = 
S'_{1} +
 \frac{1}{g_0^2} \tr \Big\{\!\mathcal F^{\mu\nu}\!a_\nu\comm{\overline{\chi}}{a_\mu}+ 
\ii\,\overline{\mathcal F}^{\mu\nu}a_\mu\comm{\chi}{a_\nu}- 
\ii\,\overline{\mathcal F}^{\mu\nu}\! a_\mu \mathcal F_{\nu\rho}a^\rho
\Big\} ~,\\
S'_{2} & \rightarrow S'_{2}(\mathcal F,
\overline{\mathcal F}) = 
S'_{2} +
\frac{1}{g_0^2} \tr \Big\{\!\!-\frac{1}{2}\epsilon_{cde}\,\lambda^c\lambda^d f^e
-f_c\,\lambda^c\eta+\ii\,f_c\,D^c\overline{\chi}+\overline{\mathcal F}_{\mu\nu} M^\mu M^\nu \Big\}~.
 \end{aligned}
\label{squartic2}
\end{equation}
Then, the full moduli action in the presence of Ramond-Ramond fluxes 
${\mathcal F}_{\mu\nu}$ and $\overline{\mathcal F}_{\mu\nu}$, and of a vacuum expectation 
value $\phi$ for the adjoint scalar of the gauge multiplet, is given by
\begin{equation}
 S'(\mathcal F, \overline{\mathcal F}, \phi) = S'_{1}(\mathcal F,
\overline{\mathcal F})+S'_{2}(\mathcal F,\overline{\mathcal F})
+{S}'_{3}(\phi)~.
\label{smodfin}
\end{equation}
This action is still BRST exact, but with respect to a modified BRST charge $Q'$. Indeed,
taking
\begin{equation}
 \label{Q1}
\begin{aligned}
& Q' a^\mu  = M^\mu~,~~~ Q' M^\mu = \ii\comm{\chi}{a^\mu}-\ii \,
{\mathcal F}^{\mu\nu} a_\nu~, \\
& Q'\lambda_c = D_c~,~~~~ Q' D_c  =  \ii\comm{\chi}{\lambda_c}+\epsilon_{cde}\,\lambda^df^e~,\\
& Q' \overline{\chi} = -\ii\,\eta~,~~~Q'\eta  = -\comm{\chi}{\overline{\chi}}~,~~~ Q' \chi = 0~, \\
& Q' \mu = h~,~~~~~~Q' h = \ii\,\mu\,\chi-\ii\,\phi\,\mu~,\\
& Q' \mu' = h'~,~~~~~ Q' h' = \ii\,\mu'\,\chi-\ii\,\phi\,\mu'~,
\end{aligned}
\end{equation}
one can check that
\begin{equation}
 S'(\mathcal F, \overline{\mathcal F}, \phi) = Q'\,\Xi'
\label{s'q'}
\end{equation}
where
\begin{equation}
 \Xi' = \Xi +\frac{1}{g_0^2} \tr \Big\{ \ii\,f_c\,\lambda^c\overline{\chi}+
\overline{\mathcal F}_{\mu\nu}a^\mu M^\nu
\Big\}
\label{xi'}
\end{equation}
with $\Xi$ defined in (\ref{Xi}). The deformed BRST charge $Q'$ is
nilpotent up to (infinitesimal) transformations of all the symmetry groups of the system; 
indeed we have
\begin{equation}
 Q'^2 \, \bullet = T_{\mathrm{SO}(k)}(\chi) \bullet\,
-\,T_{\mathrm{SU}(N)}(\phi)\bullet\, + T_{\mathrm{SU}(2)\times \mathrm{SU}(2)'}(\cF)\bullet
~,
\label{Q'2}
\end{equation}
where $T_{\mathrm{SO}(k)}(\chi)$, $T_{\mathrm{SU}(N)}(\phi)$ and $T_{\mathrm{SU}(2)\times \mathrm{SU}(2)'}(\cF)$
are infinitesimal transformations of $\mathrm{SO}(k)$, $\mathrm{SU}(N)$ 
and $\mathrm{SU}(2)\times \mathrm{SU}(2)'$, 
parametrized respectively by $\chi$, $\phi$ and $\cF$, in the
appropriate representation. Note that $\overline{\mathcal F}_{\mu\nu}$
appears only in $\Xi'$ but not in $Q'$; 
hence the variation of $S'(\mathcal F, \overline{\mathcal F}, \phi)$ 
with respect to $\overline{\mathcal F}_{\mu\nu}$ is $Q'$-exact. This fact implies
that the instanton partition function does not depend on $\overline{\mathcal F}_{\mu\nu}$,
which can therefore be set to the most convenient value for the calculations. For later purposes
it is useful to rewrite the moduli action in the following more explicit way
\begin{eqnarray}
  S'(\mathcal F, \overline{\mathcal F}, \phi)=
\frac{1}{g_0^2}\!\!\!\!&&\!\!\!\!\tr \Big\{\eta\comm{a_\mu}{M^\mu}+
\lambda^c\comm{a^\mu}{M^\nu}\bar\eta^c_{\mu\nu}
-\frac{\ii}{2}\eta\comm{\chi}{\eta} 
-\ii\,M^\mu\comm{\overline{\chi}}{M_\mu}\nonumber\\
&-&\frac{1}{2}D_c\,\bar\eta^c_{\mu\nu}
\comm{a^\mu}{a^\nu}-\comm{a_\mu}{\overline{\chi}}\comm{a^\mu}{\chi}
+\frac{1}{2}\comm{\overline{\chi}}{\chi}\comm{\overline{\chi}}{\chi}
+\cF^{\mu\nu}a_\nu\comm{\overline{\chi}}{a_\mu}\nonumber\\
&-&\frac{1}{2}\lambda_c\,{Q'}^2\lambda^c
+\frac{1}{2}D_c\,D^c-\mu^T{Q'}^2\mu' +h^Th'-f_c\,\lambda^c\eta\nonumber\\
&+&\ii\,f_c\,D^c\,\overline{\chi}+\overline{\cF}^{\mu\nu}a_\mu\,{Q'}^2a_\nu +\overline{\cF}^{\mu\nu}M_\mu M_\nu\Big\}~.
\label{S'fin}
 \end{eqnarray}
To this action we should add the classical part
\begin{equation}
 S_{\mathrm{cl}}= -2\pi\ii\tau\,k = \frac{2\pi\ii}{g_s}\,k
\label{sclass}
\end{equation}
which represents the topological normalization of the pure D(--1) disk amplitude with 
multiplicity $k$ and no moduli insertions \cite{Polchinski:1994fq,Billo:2002hm}. If a
non-zero vacuum expectation value for the Ramond-Ramond scalar $C_0$ is present, $\tau$
is promoted to the usual combination $\tau=C_0+\frac{\ii}{g_s}$.

\section{Non-perturbative effective action from stringy instantons}
\label{sec:stringyprepot}
To obtain the non-perturbative contributions to the D3 brane effective action induced by the
stringy instantons, we need to compute the partition function%
\footnote{Here, for simplicity we have omitted the exponential of minus the classical
instanton action, $\ee^{2\pi\ii\tau k}$; we will restore these factors later on.} 
\begin{equation}
 Z_k= \cN_k \int d\cM_k \,\ee^{-S'(\mathcal F, \overline{\mathcal F}, \phi)}
\label{partfunc}
\end{equation}
where $\cN_k$ is a normalization that contains also the appropriate power of the
scale factor needed to compensate for the dimensions of the moduli measure $d\cM_k$. 
For $N=2$, which is the case we will consider in detail, the
normalization $\cN_k$ is simply a numerical factor because in this case the moduli
measure is dimensionless (see \eq{dimmeas1}). We now evaluate the integrals in (\ref{partfunc})
in the semiclassical approximation, which due to the BRST structure of the instanton action
actually turns out to be exact. One way to see this is to rescale the BRST doublets
in the following way \cite{Billo:2009di}
\begin{equation}
\begin{aligned}
 &(a_\mu,M_\mu)\to\frac{1}{x}\,(a_\mu,M_\mu)~,~~(\overline{\chi},\eta)\to \frac{1}{x}\,
(\overline{\chi},\eta)~,\\
 &(\lambda_c,D_c)\to{x}^2\,(\lambda_c,D_c)~,~~(\mu,h)\to {x}^2\,(\mu,h)~,~~
(\mu',h')\to{x}^2\,(\mu',h')~,
\end{aligned}
\label{rescalings} 
\end{equation}
and the anti-holomorphic background as
\begin{equation}
 \overline{\cF}_{\mu\nu}\to z\,\overline{\cF}_{\mu\nu}~.
\label{rescF}
\end{equation}
The partition function $Z_k$ does not depend on $x$ and $z$; indeed $x$ only appears through
a change of integration variables which leaves the measure $d\cM_k$ invariant, while
$z$ is introduced through $\overline{\cF}_{\mu\nu}$ which only appears inside the gauge
fermion $\Xi'$ as shown in (\ref{xi'}). Thus, we can choose these parameters to simplify
as much as possible the structure of $Z_k$.
In particular, taking the limit
\begin{equation}
 x\to\infty~,~~z\to\infty~~~~\mbox{with}~~\frac{z}{x^2}\to\infty~,
\label{limit}
\end{equation}
the moduli action (\ref{S'fin}) reduces to
\begin{equation}
 \begin{aligned}
  S'(\mathcal F, \overline{\mathcal F}, \phi)=\tr \Big\{&\!\!
-\frac{s}{2}\lambda_c\,{Q'}^2\lambda^c
+\frac{s}{2}D_c\,D^c-s\,\mu^T{Q'}^2\mu' +s\, h^Th'-t\,f_c\,\lambda^c\eta\\
&+\ii\,t\,f_c\,D^c\,\overline{\chi}+u\,\overline{\cF}^{\mu\nu}a_\mu\,{Q'}^2a_\nu +u\,\overline{\cF}^{\mu\nu}M_\mu M_\nu\Big\}~+\cdots~.
 \end{aligned}
\label{S'rescaled}
\end{equation}
Here we have introduced the coupling constants
\begin{equation}
 s=\frac{x^4}{g_0^2}~,~~~t=\frac{x}{g_0^2}~,~~~u=\frac{z}{x^2\,g_0^2}~,
\end{equation}
which all tend to $\infty$ due to (\ref{limit}), and have denoted with dots the terms of the
first two lines of (\ref{S'fin}) which are subleading in this limit. The integrals over the
moduli can now be easily computed. 

To evaluate these integrals we choose the external background $\cF_{\mu\nu}$
along the Cartan directions of $\mathrm{SU}(2)\times \mathrm{SU}(2)'$, namely in
(\ref{fc}) we take 
\begin{equation}
 f_c= f\,\delta_{c3}~,~~~\bar f_c = \bar f\,\delta_{c3}~,
\label{fc3}
\end{equation}
so that
\begin{equation}
 \mathcal F = -\frac{\ii}{2}\,\bar f\,\eta^3 -\frac{\ii}{2}\, 
f\,\overline{\eta}^3 =-\frac{\ii}{2}\begin{pmatrix} 
      0& ~(\bar f+ f)&0&0  \cr
     -(\bar f+f)&0&0&0 \cr
      0&0&0& ~(\bar f- f)\cr
      0&0&-(\bar f- f)&0
     \end{pmatrix}~.
\label{Fcart}
\end{equation}
When the choice (\ref{fc3}) is inserted in (\ref{S'rescaled}), the fermion $\eta$ only appears in the
term proportional to $(f\,\lambda_3\,\eta)$. Thus, the integration over $\eta$ and $\lambda_3$
can be performed simultaneously producing a factor of $t\,f$, and all other terms containing
$\lambda_3$ can be neglected. On the other hand, the 
boson $\overline{\chi}$ only appears in the term proportional 
to $(f\,D_3\,\overline{\chi})$, so that the Gaussian integration over $D_3$ and 
$\overline{\chi}$ produces a factor of $1/(t\,f)$. In the end the integral over the BRST quartet
formed by $\lambda_3$, $D_3$, $\eta$ and $\overline{\chi}$ simply produces a numerical constant
which we absorb in the overall normalization factor $\cN_k$ of the instanton partition
function. 

Once this is done, we are left with the integrals over the BRST pairs $(a_\mu, M_\mu)$,
$(\mu,h)$, $(\mu',h')$ and $(\lambda_{\hat c},D_{\hat c})$ with $\hat c=1,2$, plus of course
the integral over $\chi$. The integrals over the BRST pairs are all Gaussian in the
semiclassical limit we are considering, and can be easily performed yielding
\begin{equation}
\begin{aligned}
 &\int (d\lambda_{\hat c} dD_{\hat c})~\ee^{\tr \{\frac{s}{2}\lambda_{\hat c}\,{Q'}^2\lambda^{\hat c}
-\frac{s}{2}D_{\hat c}\,D^{\hat c}\}}\,\times\,\int (d\mu dh)\, (d\mu'dh')~
\ee^{\tr \{s\,\mu^T{Q'}^2\mu' -s\, h^Th'\}}\\
&\times\,\int (da_\mu dM_\mu)~\ee^{-\tr \{
u\,\overline{\cF}^{\mu\nu}a_\mu\,{Q'}^2a_\nu +u\,\overline{\cF}^{\mu\nu}M_\mu M_\nu\}}
~\sim~ \cP(\chi)\,\times\,\cR(\chi)\,\times\,\frac{1}{\cQ(\chi)}~.
\end{aligned}
\label{int}
\end{equation}
In the last step we have defined
\begin{subequations}
\begin{align}
 \cP(\chi)&\equiv\Pf_{\big(\Yasymm,\mathbf{1},\,(\mathbf{1},\mathbf{3})'\big)}\big(Q'^2\big)~,
\label{pchi0}\\
\cR(\chi)&\equiv
\mathrm{det}_{\big(\Yfund,\,\mathbf{\overline{N}},\,(\mathbf{1},\mathbf{1})\big)}\big(Q'^2\big)~,
\label{rchi0}\\
\cQ(\chi)&\equiv
\mathrm{det}^{1/2}_{\big(\Ysymm,\,\mathbf{1},\,(\mathbf{2},\mathbf{2})\big)}\big(Q'^2\big)~,
\label{qchi0}
\end{align}
\label{PRQ}
\end{subequations}
where the labels on the Pfaffian and determinants specify the representations on which $Q'^2$
acts%
\footnote{In the first line of \eq{PRQ}, $(\mathbf{1},\mathbf{3})'$ means that the component of
the BRST pair $(\lambda_c,D_c)$ along the null weight must not be considered, since it has
been already integrated out with the quartet mechanism.}, 
and neglected all numerical factors that are absorbed in the overall normalization.  
Thus, the $k$-instanton partition function 
is given in terms of the (super) determinant of $Q'^2$ evaluated
at the fixed points of $Q'$ in agreement with the localization formulas 
\cite{Nekrasov:2002qd,Bruzzo:2002xf,Bruzzo:2003rw}, and can
be expressed in the following form 
\begin{equation}
 Z_k= \cN_k\,\int \Big\{\frac{d\chi}{2\pi\ii}\Big\}\,\frac{\cP(\chi)\,\cR(\chi)}{\cQ(\chi)}
~.
\label{zk1}
\end{equation}
Notice that, as we anticipated above, 
the result does not depend on the anti-holomorphic background $\overline{\cF}_{\mu\nu}$, nor
on the coupling constants $s$, $t$ and $u$.

Since the integrand in (\ref{zk1}) is singular when 
the denominator $\cQ(\chi)$ vanishes and tends to one when
$\chi\to\infty$, the integral over $\chi$ is naively divergent and
must be suitably defined to make sense. Here we follow exactly the same prescription of Ref.~\cite{Moore:1998et}, which has already been tested for the stringy instanton calculus in 
several explicit examples \cite{Billo:2009di}-\nocite{Fucito:2009rs,Billo:2010bd}\cite{Billo:2010mg}. 
In particular, we cure the singularities along the 
integration path by giving the zeroes of $\cQ(\chi)$ a small positive imaginary part moving them 
in the upper-half complex plane, and regulate the divergence at infinity by interpreting the
$\chi$-integral as a contour integral.

\subsection{Explicit results for small instanton numbers}
We will now derive the explicit expression of the partition function for low instanton numbers
in the SU(2) theory. The case of $\mathrm{SU}(N)$ will be considered in a separate publication
\cite{progress}.

\subsubsection{$k=1$}
The 1-instanton partition function $Z_1$ is particularly simple: in fact, for $k=1$ there are no $\lambda_c$'s and no $\chi$'s, so that the factor $\cP(\chi)$ is not generated and no
contour integral has to be evaluated. Furthermore, for $k=1$ we simply have
\begin{equation}
 \begin{aligned}
  \cR(\chi) &\propto\,\det\phi~,\\
\cQ(\chi) &\propto\,\mathrm{det}^{1/2}{\cF}\,\propto\, E_1 E_2 \equiv \cE~,
 \end{aligned}
\label{RQ1}
\end{equation}
where we have defined
\begin{equation}
\label{ef}
 E_1=\frac{f+\bar f}{2}~,~~~E_2=\frac{f-\bar f}{2}~,
\end{equation}
and neglected all numerical factors. Absorbing the latter into the overall normalization, we
eventually find
\begin{equation}
 Z_1 = {\cN}_1\, \frac{\det\phi}{\cE}~.
\label{z1}
\end{equation}
Notice that the factor $1/\cE$ in the above result can be interpreted as the regulated volume of the four-dimensional $\cN=2$ superspace \cite{Nekrasov:2002qd,Billo:2006jm}, 
since for $k=1$ the moduli ${a_\mu}$ and ${M_\mu}$ are identified with the superspace coordinates.

\subsubsection{$k>1$}
In this case, in order to perform the integration over the $\chi$'s 
we exploit the $\mathrm{SO}(k)$ invariance of the
integrand in (\ref{zk1}) and, at the price of introducing a Vandermonde determinant
$\Delta(\chi)$, bring the $\chi$'s to the Cartan subalgebra, 
whose generators we denote as $H^i_{\mathrm{SO}(k)}$, {\it i.e.}
\begin{equation}
 \label{chicartan}
 \chi ~~\to~~ \vec\chi \cdot \vec H_{\mathrm{SO}(k)}~= \sum_{i=1}^{\mathrm{rank}\, \mathrm{SO}(k)}
\chi_{i} \,H^{i}_{\mathrm{SO}(k)} ~.
\end{equation}
Then the partition function becomes
\begin{equation}
\label{Zkred}
Z_k = \cN_k \int \prod_{i} \Big(\frac{d \chi_{i}}{2\pi\ii}\Big)~ \Delta(\vec\chi)\,
\frac{\cP(\vec \chi)\, \cR(\vec\chi)}{\cQ(\vec\chi)}~.
\end{equation}
Again, we have absorbed all numerical factors produced by the ``diagonalization'' of $\chi$ into a redefinition of the normalization coefficient $\cN_k$.
Furthermore, without any loss of generality we can assume that also the vacuum 
expectation value $\phi$ belongs to the Cartan direction
of $\mathrm{SU}(2)$, namely
\begin{equation}
\phi =\frac{\varphi}{2}\,\tau^3~.
\label{phisu2}
\end{equation}

Let us now consider the 2-instanton partition function. As shown in detail in Appendix \ref{app:A},
for $k=2$ we have
\begin{equation}
  \begin{aligned}
\cP(\vec \chi) & \propto\,-(E_1+E_2)~,~~~
\cR(\vec\chi) \propto\,
\big(\chi^2+\det\phi\big)^2~,\\
  \cQ(\vec\chi)& \propto\,\cE\,\prod_{A=1}^2(2\chi-E_A)(2\chi+E_A)~,~~~
\Delta(\vec\chi)=1~,
\end{aligned}
\label{ingrk2}
 \end{equation}
so that
\begin{equation}
 Z_2=-\cN_2\,\frac{E_1+E_2}{\cE}\,\int\frac{d\chi}{2\pi\ii}~
\frac{\big(\chi^2+\det\phi\big)^2}{(4\chi^2-E_1^2)(4\chi^2-E_2^2)}~.
\label{z2}
\end{equation}
As we mentioned in the previous subsection, the $\chi$-integral must be 
understood as a contour integral in the upper-half complex plane and 
the singularities at the zeroes of the denominator in
(\ref{z2}) are avoided by giving the deformation parameters $E_A$ a small positive 
imaginary part, according to the prescriptions of Ref.s~\cite{Moore:1998et}. 
In particular, we choose
\begin{equation}
 \label{imparts}
 \im E_1 > \im E_2 > \im \frac{E_1}{2} > \im \frac{E_2}{2} > 0~.
\end{equation}
Evaluating the residues, we finally obtain
\begin{equation}
 Z_{2} = \frac{\cN_2}{4\,\cE^2}\,\mathrm{det}^2\phi 
- \frac{\cN_2}{8\,\cE}\, \det\phi
- \frac{\cN_2}{64\,\cE} \big[(E_1^2+E_2^2)+\cE\big]~.
 \label{z2res}
\end{equation}

The calculation for $k=3$ proceeds in the same way, even if it is algebraically
a bit more involved. Some technical details are given in Appendix \ref{app:A}; here we simply quote the
final result, namely
\begin{equation}
 Z_{3} = \frac{\cN_3}{12\,\cE^3}\,\mathrm{det}^3\phi 
- \frac{\cN_3}{8\,\cE^2}\,\mathrm{det}^2\phi
- \frac{\cN_3}{192\,\cE^2}\big[3(E_1^2+E_2^2)-5\cE\big]\,\det\phi~.
 \label{z3}
\end{equation}
The explicit expressions for $Z_4$ and $Z_5$ can be obtained as well and they 
are given in Appendix \ref{app:A}. Since they are rather cumbersome, 
we refrain from writing them here; however, we report the terms with the highest 
power of $\cE$ in the denominator, namely
\begin{subequations}
 \begin{align}
  Z_4 &=\frac{\cN_4}{48\,\cE^4}\,\mathrm{det}^4\phi + \cdots~,\label{z4a}\\
  Z_5 &= \frac{\cN_5}{240\,\cE^5}\,\mathrm{det}^5\phi + \cdots ~,
\label{z5a}
 \end{align}
\end{subequations}
which will be useful for the subsequent calculations.

\subsection{The non-perturbative prepotential}
{From} the partition functions $Z_k$ computed above, we define
the ``grand-canonical'' instanton partition function
\begin{equation}
 \label{Ztot}
\mathcal Z =
\sum_{k=0}^\infty Z_k\, \ee^{2\pi\ii\tau k} =
\sum_{k=0}^\infty Z_k\, q^k
\end{equation}
where we have set $Z_0=1$ and $q \equiv\exp(2\pi\ii\tau)$.
To obtain the non-perturbative D3 brane effective action
induced by the stringy instantons and remove the disconnected contributions, 
we have to take the logarithm of $\mathcal Z$.
Notice that the partition functions $Z_k$ have been computed by integrating 
over {\it all} moduli, including the instanton ``center-of-mass'' 
coordinates and their superpartners playing the r\^ole of the superspace coordinates.
In the absence of the Ramond-Ramond deformations these zero-modes do not appear in the moduli action and the integration over them would diverge. In the presence of deformations, instead, this integration
yields a factor of $1/\mathcal{E}$ (as is clearly seen from the $k=1$ result (\ref{z1})), 
and thus to extract the integral over the centered moduli only, it is sufficient 
to multiply $\log \cZ$ by $\cE$. Having done so, we can promote the vacuum
expectation value $\phi$ appearing in $\cZ$ to the full fledged dynamical superfield $\Phi(x,\theta)$ and, after removing the Ramond-Ramond deformations, we finally 
obtain the non-perturbative contributions to the D3 brane effective action, namely
\begin{equation}
 \label{SeffPhi}
 S^{\mathrm{(n.p.)}} = \int d^4x\, d^4\theta\, 
F^{\mathrm{(n.p.)}}\big(\Phi(x,\theta)\big)
\end{equation}
where the ``prepotential'' $F^{\mathrm{(n.p.)}}(\Phi)$ is
\begin{equation}
 \label{prep1}
 F^{\mathrm{(n.p.)}}(\Phi) = {\mathcal{E}}\, \log \mathcal Z\Big|_{\phi\to\Phi,E_A\to 0}~.
\end{equation}
Expanding in powers of $q$, we have
\begin{equation}
 \label{Fexp}
 F^{\mathrm{(n.p.)}}(\Phi) = \sum_{k=1}^\infty F_k\, q^k~\Big|_{\phi\to\Phi,E_A\to 0}
\end{equation}
where the first few coefficients are
\begin{equation}
 \label{Fkexp}
 \begin{aligned}
  F_1 & = \cE Z_1~,\\
  F_2 & = \cE Z_2 - \frac {F_1^2}{2\cE}~,\\
  F_3 & = \cE Z_3 - \frac{F_2 F_1}{\cE} - \frac{F_1^3}{6\cE^2}~,\\
  F_4 & = \cE Z_4 - \frac{F_3 F_1}{\cE} - \frac{F_2^2}{2\cE} -
          \frac{F_2 F_1^2}{2\cE^2} - \frac{F_1^4}{24\cE^3}~,\\
  F_5 & = \cE Z_5 - \frac{F_4 F_1}{\cE} - \frac{F_3 F_2}{\cE} - 
           \frac{F_3 F_1^2}{2\cE^2} - \frac{F_2^2 F_1}{2\cE^2} -
           \frac{F_2 F_1^3}{6\cE^3} - \frac{F_1^5}{120\cE^4}~.
 \end{aligned}
\end{equation}
The prepotential $F^{\mathrm{(n.p.)}}(\Phi)$ must be well-defined when the closed-string deformations
are turned off, and thus all coefficients $F_k$ must be finite in the limit $E_A\to 0$.
{From} (\ref{Fkexp}) and the expressions of $Z_k$ we derived in the previous subsection,
we see that the $F_k$'s contain singular terms diverging as $1/\cE^{k-1},~1/\cE^{k-2},\cdots$
for $E_A\to 0$.
Imposing the cancellation of the most divergent terms of $F_k$ fixes the overall normalization 
$\cN_k$ but, once this choice is made, no freedom is left and all the remaining divergences 
must disappear. Verifying that this happens is a very strong check on our results and
the consistency of the whole procedure.

For $k=1$, from \eq{z1} we have directly 
\begin{equation}
 \label{F1is}
  F_1 = \cN_1\,\det\phi~.
\end{equation}
This is the same result obtained in \cite{Argurio:2007vqa}.
For $k=2$, from Eq.s (\ref{Fkexp}) and (\ref{z2res}) we find
\begin{equation}
 \label{mdivF2}
 F_2 = \left(\frac{\cN_2}{4} - \frac{\cN_1^2}{2}\right)\frac{\mathrm{det}^2\phi}{\cE}
- \frac{\cN_2}{8}\, \det\phi
- \frac{\cN_2}{64} \big[(E_1^2+E_2^2)+\cE\big]~. 
\end{equation}
If we choose
\begin{equation}
 \label{n2sol}
 \cN_2 = 2 \cN_1^2~,
\end{equation}
the most divergent term disappears, and we are left with
\begin{equation}
 \label{F2sol}
 F_2= 
- \frac{\cN_1^2}{4}\, \det\phi
- \frac{\cN_1^2}{32} \big[(E_1^2+E_2^2)+\cE\big]
\end{equation}
which is indeed finite when $E_A\to 0$.
We then proceed in the same way at the next order, $k=3$. Using \eq{z3} and inserting the
above expressions for $F_1$ and $F_2$ in (\ref{Fkexp}), we find
\begin{equation}
\label{F3div}
F_3  = 
\left(\frac{\cN_3}{12} -\frac{\cN_1^3 }{6}\right)\frac{\mathrm{det}^3\phi}{\mathcal{E}^2}  
+  \ldots ~,
\end{equation}
so that we have to choose
\begin{equation}
 \label{n3sol}
 \cN_3 = 2 \cN_1^3~.
\end{equation}
Once this is done, all other divergences in $F_3$ cancel and we are simply left with
\begin{equation}
 F_3 = \frac{\cN_1^3}{12} \,\det\phi~.
\label{F3is}
\end{equation}
For $k=4$, we use the partition function $Z_4$ given in Appendix \ref{app:A} and again require
the cancellation of most divergent term in the resulting expression for $F_4$ following from \eq{Fkexp}. This fixes $\cN_4 = 2 \cN_1^4$. Using this, we then find
\begin{equation}
 \label{f4res}
 F_4 = -\frac{\cN_1^4}{32} \,\det\phi - \frac{\cN_1^4}{256} \big[(E_1^2+E_2^2)+\cE\big]~,
\end{equation}
which has a finite limit when $E_A\to 0$.
In the case $k=5$, having computed $Z_5$ along the lines described in Appendix \ref{app:A}, the cancellation of the highest divergence in $F_5$ leads to $\cN_5 = 2 \cN_1^5$, after which we get 
\begin{equation}
\label{f5res}
 F_5 = \frac{\cN_1^5}{80} \,\det\phi~.
\end{equation}

Making the replacement $\phi\to\Phi(x,\theta)$ and taking the limit $E_A\to 0$ in the above results, we finally obtain from the non-perturbative prepotential of the SU(2) gauge theory
induced by the stringy instantons.
Up to instanton number $k=5$, our findings are summarized in
\begin{equation}
 \label{Fupto5}
  \begin{aligned}
   F^{\mathrm{(n.p.)}}(\Phi)  = -\Tr\Phi^2\,\Big(\frac{\cN_1}{2}\,q - \frac{\cN_1^2}{8}\,q^2 +
\frac{\cN_1^3}{24}\,q^3-\frac{\cN_1^4}{64}\,q^4+\frac{\cN_1^5}{160}\,q^5 \ldots\Big)~,
  \end{aligned}
\end{equation}
where we made use of the relation $\det\phi=-\frac{1}{2}\,\Tr\phi^2$ which easily
follows from \eq{phisu2}.

\section{Summary and conclusions}
\label{sec:concl}
The detailed analysis presented in the previous sections shows that 
the stringy instantons have the right content of zero-modes to produce 
non-perturbative terms in the $\cN=2$
$\mathrm{SU}(N)$ theories in four dimensions realized with fractional D3 branes 
in a $\mathbb{C}^3/\mathbb{Z}_3$ orientifold.
For the SU(2) model such terms have been explicitly computed using localization methods up 
to instanton number $k=5$, and have been shown to provide non-perturbative corrections 
to the effective prepotential of the theory.

It is worth to remark that these results are unconventional from a purely field-theory
point of view, but are quite natural in the stringy approach to the instanton calculus. In fact,
the exotic instantons in our model are fractional D(--1) branes that occupy 
quiver nodes where no D3 branes are present but, apart from this feature, they are completely
standard D-instantons, possessing their ``own life'' independently of the existence
of an underlying gauge theory. What is non-standard, however, is the content of their moduli space: 
indeed, the charged zero modes corresponding to the mixed open strings stretching between the stringy instantons and the gauge branes are only fermionic, and the neutral zero-modes corresponding
to open strings with both end-points on the stringy instantons are in representations of
orthogonal groups even if the gauge groups are unitary.
This is to be contrasted with what happens for the ordinary instantons in theories with
unitary gauge groups, where the charged zero-modes are both bosonic and fermionic and 
the neutral zero-modes fall into representations of unitary groups if the gauge group is unitary. 
These differences result in a different structure of the moduli integral and 
in a different scaling dimension
of the integration measure on moduli space. For the $\mathrm{SU}(2)$ model the integration measure
turns out to be dimensionless (see \eq{dimmeas1}) and thus the prepotential of the theory can receive
contributions from exotic configurations with any instanton number%
\footnote{This is similar to what happens with ordinary instantons in the $\cN=2$ SU(2) 
gauge theory with four fundamental flavors.}.  Notice that in this $\mathrm{SU}(2)$ model 
the supersymmetry is enhanced at tree-level 
from $\cN=2$ to $\cN=4$, because the SU(2) symmetric representation 
in which the hypermultiplet transforms is equivalent to the adjoint representation. 
Therefore, this model can be regarded as a non-conventional
realization of an $\cN=4$ SU(2) super Yang-Mills theory in four dimensions. 
As is well known, the usual gauge instantons in this case do not contribute to the 
(quadratic) effective action; on the contrary, as we have explicitly shown, the stringy instantons 
do. Furthermore, since they only correct the prepotential, the above supersymmetry enhancement is lost at the non-perturbative level. 
{From} our results up to instanton number $k=5$ 
(see \eq{Fupto5}), it is very natural to conjecture that 
the stringy instanton series of the $\mathrm{SU}(2)$ theory can be resummed into 
 \begin{equation}
 \label{Fresummed}
   F^{\mathrm{(n.p.)}}(\Phi)  = -\Tr\Phi^2\,\log\Big(1+\frac{\cN_1}{2}\,q\Big)
\end{equation}
where the non-vanishing constant $\cN_1$ can be fixed by a careful analysis of the normalization
of the 1-instanton partition function.
This seems to suggest that the stringy instantons induce a non-perturbative redefinition of the
gauge coupling constant of logarithmic type, so that the SU(2) prepotential appears classical 
in terms of the new coupling.  It would be interesting to understand whether this non-perturbative redefinition has some deeper meaning.

\vskip 1cm

\noindent {\large {\bf Acknowledgments}}
\vskip 0.2cm
We thank Marialuisa Frau and Igor Pesando for many useful discussions and especially Marco Bill\`o
for sharing with us his insight and for his help with the Mathematica programs.

\vfill\eject

\appendix

\section{Details on the D-instanton computations}
\label{app:A} 
The expressions of the functions $\cP(\chi)$, $\cR(\chi)$, $\cQ(\chi)$ defined in 
(\ref{PRQ}) and appearing in the integrand of the instanton partition function 
(\ref{zk1}), can be expressed in terms of the weights of the
relevant representations of the instantonic symmetry group $\mathrm{SO}(k)$,
of the twisted Lorentz group $\mathrm{SU}(2)\times\mathrm{SU}(2)'$
and of the gauge group SU(2) .
For convenience we recall the form of these weight vectors.
\paragraph{Weight sets of SO$(2n+1)$:}
This group has rank $n$. If we denote by $\ve{i}$ the versors in the 
$\mathbb{R}^n$ weight space, then
\begin{itemize}
 \item the set of the $2n+1$ weights $\vec\pi$ of the vector representation is 
 given by
 \begin{equation}
  \label{setvec1}
  \pm\ve i~,~~~~ \vec 0~~\mbox{with multiplicity $1$}~;
 \end{equation}
 \item the set of $n(2n+1)$ weights $\vec\rho$ of the adjoint representation (corresponding
to the two-index antisymmetric tensor) is the following:
\begin{equation}
 \label{setrho1}
 \pm \ve i \pm \ve j~(i < j)~,~~~~
 \pm \ve i~,~~~~
 \vec 0~~\mbox{with multiplicity $n$}~;
\end{equation}
\item the $(n+1)(2n+1)$ weights $\vec\sigma$ of the two-index symmetric tensor%
\footnote{In fact, this is not an irreducible representation: it decomposes 
into the $(n+1)(2n+1)-1$ traceless symmetric tensor plus a singlet. One of the
$\vec 0$ weights corresponds to the singlet.} are
\begin{equation}
 \label{setsymm1}
 \pm \ve i \pm \ve j~(i < j)~,~~~~
 \pm \ve i~,~~~~
 \pm 2 \ve i~,~~~~
 \vec 0~~\mbox{with multiplicity $n+1$}~.
\end{equation} 
\end{itemize}

\paragraph{Weight sets of SO$(2n)$:}
This group has rank $n$.
If we denote by $\ve{i}$ the versors in the 
$\mathbb{R}^n$ weight space, then
\begin{itemize}
 \item the set of the $2n$ weights $\vec\pi$ of the vector representation is 
 given by
 \begin{equation}
  \label{setvec}
  \pm\ve i~;
 \end{equation}
 \item the set of $n(2n-1)$ weights $\vec\rho$ of the two-index antisymmetric 
tensor is the following:
\begin{equation}
 \label{setrho}
 \pm \ve i \pm \ve j~(i < j)~,~~~~
 \vec 0~~\mbox{with multiplicity $n$}~;
\end{equation}
\item the $n(2n+1)$ weights $\vec\sigma$ of the two-index symmetric tensor%
\footnote{Again, this is not an irreducible representation, since it 
contains a singlet.} are
\begin{equation}
 \label{setsymm}
 \pm \ve i \pm \ve j~(i < j)~,~~~~
 \pm 2 \ve i~,~~~~
 \vec 0~~\mbox{with multiplicity $n$}~.
\end{equation}
\end{itemize}

\paragraph{Weight sets of SU$(2)\times$SU$(2)'$:} The relevant representations of the twisted
Lorentz group are the $(\mathbf{1,3})$ in which the BRST pair $(\lambda_c,D_c)$ transforms, and the
$(\mathbf{2,2})$ in which the BRST pair $(a_\mu,M_\mu)$ transforms.
\begin{itemize}
 \item the weights $\vec\alpha$ of the $(\mathbf{1,3})$ representation are given by the
following two-component vectors
\begin{equation}
 (0,\pm1)~,~~~~(0,0)~.
\label{alpha1}
\end{equation}
In our conventions, the weight $(0,+1)$ is considered to be positive.
\item the weights $\vec\beta$ of the $(\mathbf{2,2})$ representation are given by the
following two-component vectors
\begin{equation}
\big(\!\pm 1/2,\pm 1/2\big)~.
\label{beta1}
\end{equation}
The weights $\big(\!\pm 1/2,+1/2\big)$ are considered positive in our conventions.
\end{itemize}

\paragraph{Weight sets of SU$(2)$:} In this case, the only relevant SU(2) representation
that occurs in our analysis is the fundamental one, for which the two weights $\vec \gamma$ is
simply given by $\pm1/2$. 

To evaluate the moduli integral and obtain the instanton partition function it is convenient
to align the vacuum expectation value $\phi$ of the chiral multiplet along the Cartan direction of SU(2), and the external Ramond-Ramond background $\cF$ along the Cartan directions of
$\mathrm{SU}(2)\times\mathrm{SU}(2)'$, namely
\begin{equation}
\phi = \vec{\phi} \cdot \vec{H}_{\mathrm{SU}(2)}~~~\mbox{and}~~~
\cF = \vec{f} \cdot \vec{H}_{\mathrm{SU}(2)\times\mathrm{SU}(2)'}
\label{phiF_cart}~.
\end{equation}
Comparing with Eq.s (\ref{Fcart}) and (\ref{phisu2}), we see that
\begin{equation}
 \vec\phi = \varphi~~~\mbox{and}~~~\vec f = (\bar f, f)~.
\label{fphi}
\end{equation}
Furthermore, exploiting the $\mathrm{SO}(k)$ invariance, we arrange the $\chi$ moduli along the
Cartan directions, namely
\begin{equation} 
\label{chi_cart}
 \chi \rightarrow \vec{\chi} \cdot \vec{H}_{\mathrm{SO}(k)} = \sum_{i=1}^n \chi_i H^i_{\mathrm{SO}(k)}~,
\end{equation}
at the price of introducing in the integrand a Vandermonde determinant given by
\begin{equation}
 \label{vanderm}
\Delta(\vec\chi) =
\prod_{\vec\rho\not=\vec 0}\vec\chi\cdot \vec\rho 
=\left\{
 \begin{aligned}
&  \prod_{i<j} 
\big(\chi_i^2 -\chi_j^2\big)^2 & \mbox{for } k=2n~,\\
 &(-1)^n\prod_{i=1}^n \chi_i^2\prod_{j<\ell}\big(\chi_j^2-\chi_\ell^2\big) ^2& \mbox{for } k=2n+1~.
 \end{aligned}
 \right.
\end{equation}

With all these definitions at hand, we can now give the explicit expressions for the functions
$\cP(\chi)$, $\cR(\chi)$, $\cQ(\chi)$. {From} \eq{pchi0}, we have 
\begin{equation}
 \label{pchi}
\begin{aligned}
 \cP(\vec\chi) &= \prod_{\vec\rho}\prod_{\vec\alpha}^+\left(\vec\chi\cdot
\vec\rho -\vec f\cdot \vec\alpha\right)\\
&= \left\{
 \begin{aligned}
 & (-f)^n
\prod_{i<j}^n\left[(\chi_i+\chi_j)^2-f^2\right]\left[(\chi_i-\chi_j)^2-f^2\right]& \mbox{for } k=2n~,\\
 &f^n\prod_{i}^n\big(\chi_i^2-f^2\big)
\prod_{j<\ell}^n
\left[(\chi_j+\chi_\ell)^2-f^2\right]\left[(\chi_j-\chi_\ell)^2-f^2\right]
& \mbox{for } k=2n+1~.
 \end{aligned}
 \right.
\end{aligned}
\end{equation}
where the product over $\vec\alpha$ is limited to the positive weight $(0,+1)$. 
This is the meaning of the superscript $+$ appearing above. 
{From} \eq{rchi0}, we have 
\begin{equation}
 \label{rchi}
\begin{aligned}
 \cR(\vec\chi) = \prod_{\vec\pi} \prod_{\vec\gamma}\left(\vec\chi\cdot
\vec\pi -\vec\phi\cdot \vec\gamma\right) =\left\{
 \begin{aligned}
 & \prod_{i=1}^n \big(\chi_i^2+\det\phi\big)^2 & \mbox{for } k=2n~,\\
 & \det\phi\prod_{i=1}^n \big(\chi_i^2+\det\phi\big)^2 & \mbox{for } k=2n+1~.
 \end{aligned}
 \right.
\end{aligned}
\end{equation}
and finally from \eq{qchi0}, we have
\begin{equation}
 \label{qchi}
\begin{aligned}
 \cQ(\vec\chi) &= \prod_{\vec\sigma} \prod_{\vec\beta}^+\left(\vec\chi\cdot
\vec\sigma -\vec f\cdot \vec\beta\right)\\
&=\left\{
 \begin{aligned}
 &\cE^n\prod_{A=1}^2 \prod_{i=1}^n
\big(4\chi_i^2-E_A^2\big)
\prod_{j<\ell}\left[(\chi_j+\chi_\ell)^2-E_A^2\right]\left[(\chi_j-\chi_\ell)^2-E_A^2\right]& \mbox{for } k=2n~,\\
 &\cE^{n+1}\prod_{A=1}^2 \Bigg\{\prod_{i=1}^n
\big(\chi_i^2-E_A^2\big)\big(4\chi_i^2-E_A^2\big)~\times\\
&\hspace{40pt}\times
\prod_{j<\ell}\left[(\chi_j+\chi_\ell)^2-E_A^2\right]\left[(\chi_j-\chi_\ell)^2-E_A^2\right]\Bigg\}
& \hspace{-30pt}\mbox{for } k=2n+1~,
 \end{aligned}
 \right.
\end{aligned}
\end{equation}
where again the product over $\vec\beta$ is limited to the positive weights.

Using these definitions and recalling from \eq{ef} that $f=(E_1+E_2)$, it is easy to find that at instanton number $k=2$ the partition function
(\ref{Zkred}) reads
\begin{equation}
 Z_2=-\cN_2\,\frac{E_1+E_2}{\cE}\,\int\frac{d\chi}{2\pi\ii}~
\frac{\big(\chi^2+\det\phi\big)^2}{(4\chi^2-E_1^2)(4\chi^2-E_2^2)}~.
\label{z2app}
\end{equation}
as reported in \eq{z2} of the main text. Evaluating the $\chi$ integral as a contour integral in the
upper half complex plane with the pole prescription (\ref{imparts}), and summing the residues at
$\chi=E_A$ and $\chi=E_A/2$ for $A=1,2$, we eventually find the result given in \eq{z2res}.
Proceeding in a similar way, at instanton number $k=3$ we find
\begin{equation}
 Z_3= -\cN_3\,\frac{\det\phi\,(E_1+E_2)}{\cE^2}\int\frac{d\chi}{2\pi\ii}~
\frac{\big(\chi^2-(E_1+E_2)^2\big)\big(\chi^2+\det\phi\big)^2}{
(\chi^2-E_1^2)(\chi^2-E_2^2)(4\chi^2-E_1^2)(4\chi^2-E_2^2)}~,
\label{z3app}
\end{equation}
from which the result given in \eq{z3} follows.

We conclude by giving the explicit expressions of the instanton partition functions at $k=4$
and $k=5$. They are
\begin{eqnarray}
 Z_{4} &= &\frac{\cN_4}{48\,\cE^4}\,\mathrm{det}^4\phi -\frac{\cN_4}{16\,\cE^3}\,\mathrm{det}^3\phi
 - \frac{\cN_4}{384\,\cE^3}\big[3(E_1^2+E_2^2)-19\cE\big]\mathrm{det}^2\phi
\label{z4app}\\
&&+\, \frac{\cN_4}{256\,\cE^2}\big[(E_1^2+E_2^2)-3\cE\big]\det\phi
+\frac{\cN_4}{4096\,\cE^2}\big[(E_1^2+E_2^2)-7\cE\big]\big[(E_1^2+E_2^2)+\cE\big]~,
 \nonumber
\end{eqnarray}
and
\begin{eqnarray}
 Z_{5} &= &\frac{\cN_5}{240\,\cE^5}\,\mathrm{det}^5\phi -\frac{\cN_4}{48\,\cE^4}\,\mathrm{det}^4\phi
 - \frac{\cN_5}{384\,\cE^4}\big[(E_1^2+E_2^2)-13\cE\big]\mathrm{det}^3\phi\nonumber\\
&&+\frac{\cN_5}{768\,\cE^3}\big[3(E_1^2+E_2^2)-17\cE\big]\mathrm{det}^2\phi
\label{z5app}\\
&&+\, \frac{\cN_5}{61440\,\cE^3}\big[15(E_1^4+E_2^4)-170 (E_1^2+E_2^2)+299\cE^2\big]\det\phi
~. \nonumber
\end{eqnarray}

\providecommand{\href}[2]{#2}\begingroup\raggedright\endgroup

\end{document}